\documentclass[runningheads]{llncs}
\usepackage[hyphens]{url}
\usepackage{hyperref}
\usepackage[hyphenbreaks]{breakurl}

\usepackage[T1]{fontenc}
\usepackage{graphicx}
\makeatletter
\@ifundefined{lhs2tex.lhs2tex.sty.read}%
  {\@namedef{lhs2tex.lhs2tex.sty.read}{}%
   \newcommand\SkipToFmtEnd{}%
   \newcommand\EndFmtInput{}%
   \long\def\SkipToFmtEnd#1\EndFmtInput{}%
  }\SkipToFmtEnd

\newcommand\ReadOnlyOnce[1]{\@ifundefined{#1}{\@namedef{#1}{}}\SkipToFmtEnd}
\usepackage{amstext}
\usepackage{amssymb}
\usepackage{stmaryrd}
\DeclareFontFamily{OT1}{cmtex}{}
\DeclareFontShape{OT1}{cmtex}{m}{n}
  {<5><6><7><8>cmtex8
   <9>cmtex9
   <10><10.95><12><14.4><17.28><20.74><24.88>cmtex10}{}
\DeclareFontShape{OT1}{cmtex}{m}{it}
  {<-> ssub * cmtt/m/it}{}

\DeclareFontShape{OT1}{cmtt}{bx}{n}
  {<5><6><7><8>cmtt8
   <9>cmbtt9
   <10><10.95><12><14.4><17.28><20.74><24.88>cmbtt10}{}
\DeclareFontShape{OT1}{cmtex}{bx}{n}
  {<-> ssub * cmtt/bx/n}{}

\newcommand{\anonymous}{\kern0.06em \vbox{\hrule\@width.5em}}
\newcommand{\plus}{\mathbin{+\!\!\!+}}

\newcommand{\sequ}{\mathbin{>\!\!\!>}}

\usepackage{polytable}

\@ifundefined{mathindent}%
  {\newdimen\mathindent\mathindent\leftmargini}%
  {}%

\def\resethooks{%
  \global\let\SaveRestoreHook\empty
  \global\let\ColumnHook\empty}
\newcommand*{\savecolumns}[1][default]%
  {\g@addto@macro\SaveRestoreHook{\savecolumns[#1]}}
\newcommand*{\restorecolumns}[1][default]%
  {\g@addto@macro\SaveRestoreHook{\restorecolumns[#1]}}
\newcommand*{\aligncolumn}[2]%
  {\g@addto@macro\ColumnHook{\column{#1}{#2}}}

\resethooks

\newcommand{\onelinecommentchars}{\quad-{}- }
\newcommand{\commentbeginchars}{\enskip\{-}
\newcommand{\commentendchars}{-\}\enskip}

\newcommand{\visiblecomments}{%
  \let\onelinecomment=\onelinecommentchars
  \let\commentbegin=\commentbeginchars
  \let\commentend=\commentendchars}

\newcommand{\invisiblecomments}{%
  \let\onelinecomment=\empty
  \let\commentbegin=\empty
  \let\commentend=\empty}

\visiblecomments

\newlength{\blanklineskip}
\newcommand{\hsindent}[1]{\quad}%
\let\hspre\empty
\let\hspost\empty

\EndFmtInput
\makeatother
\ReadOnlyOnce{polycode.fmt}%
\makeatletter

\newcommand{\hsnewpar}[1]%
  {{\parskip=0pt\parindent=0pt\par\vskip #1\noindent}}

\newcommand{\hscodestyle}{}

\newcommand{\sethscode}[1]%
  {\expandafter\let\expandafter\hscode\csname #1\endcsname
   \expandafter\let\expandafter\endhscode\csname end#1\endcsname}

  {\par\noindent
   \advance\leftskip\mathindent
   \hscodestyle
   \let\\=\@normalcr
   \let\hspre\(\let\hspost\)%
   \pboxed}%
  {\endpboxed\)%
   \par\noindent
   \ignorespacesafterend}

  {\hsnewpar\abovedisplayskip
   \advance\leftskip\mathindent
   \hscodestyle
   \let\hspre\(\let\hspost\)%
   \pboxed}%
  {\endpboxed%
   \hsnewpar\belowdisplayskip
   \ignorespacesafterend}

  {\hsnewpar\abovedisplayskip
   \advance\leftskip\mathindent
   \hscodestyle
   \let\\=\@normalcr
   \(\pboxed}%
  {\endpboxed\)%
   \hsnewpar\belowdisplayskip
   \ignorespacesafterend}

\newcommand{\plainhs}{\sethscode{plainhscode}}

\plainhs

  {\hsnewpar\abovedisplayskip
   \advance\leftskip\mathindent
   \hscodestyle
   \let\\=\@normalcr
   \(\parray}%
  {\endparray\)%
   \hsnewpar\belowdisplayskip
   \ignorespacesafterend}

  {\parray}{\endparray}

  {\(\parray}{\endparray\)}

\def\codeframewidth{\arrayrulewidth}
\RequirePackage{calc}

  {\parskip=\abovedisplayskip\par\noindent
   \hscodestyle
   \arrayrulewidth=\codeframewidth
   \tabular{@{}|p{\linewidth-2\arraycolsep-2\arrayrulewidth-2pt}|@{}}%
   \hline\framedhslinecorrect\\{-1.5ex}%
   \let\endoflinesave=\\
   \let\\=\@normalcr
   \(\pboxed}%
  {\endpboxed\)%
   \framedhslinecorrect\endoflinesave{.5ex}\hline
   \endtabular
   \parskip=\belowdisplayskip\par\noindent
   \ignorespacesafterend}

\newcommand{\framedhslinecorrect}[2]%
  {#1[#2]}

  {\(\def\column##1##2{}%
   \let\>\undefined\let\<\undefined\let\\\undefined
   \newcommand\>[1][]{}\newcommand\<[1][]{}\newcommand\\[1][]{}%
   \def\fromto##1##2##3{##3}%
   }{\) }%

  {\let\orighscode=\hscode
   \let\origendhscode=\endhscode
   \def\endhscode{\def\hscode{\endgroup\def\@currenvir{hscode}\\}\begingroup}
   \orighscode\def\hscode{\endgroup\def\@currenvir{hscode}}}%
  {\origendhscode
   \global\let\hscode=\orighscode
   \global\let\endhscode=\origendhscode}%

\makeatother
\EndFmtInput
\usepackage{xcolor}

\definecolor{hsgold2}{RGB/cmyk}{177,149,90/0,0.16,0.49,0.3}
\definecolor{hsgold3}{RGB/cmyk}{190,106,13/0,0.44,0.93,0.25}
\definecolor{hsblue3}{RGB/cmyk}{0,33,132/1,0.65,0,0.35}
\definecolor{hsblue4}{RGB/cmyk}{97,108,132/0.26,0.18,0,0.48}
\definecolor{hsblue5}{RGB/cmyk}{34,50,68/0.5,0.26,0,0.73}
\definecolor{hsred2}{RGB/cmyk}{191,121,103/0,0.4,0.49,0.23}
\definecolor{hsred3}{RGB/cmyk}{171,72,46/0,0.58,0.73,0.33}

\colorlet{HSBLUE3}{hsblue3}

\newcommand*{\mathcolor}{}
\def\mathcolor#1#{\mathcoloraux{#1}}
\newcommand*{\mathcoloraux}[3]{%
  \protect\leavevmode
  \begingroup
    \color#1{#2}#3%
  \endgroup
}

\newcommand{\HSKeyword}[1]{\mathcolor{hsgold3}{\textbf{#1}}}

\newcommand{\HSSpecial}[1]{\mathcolor{hsblue4}{#1}}
\newcommand{\HSSym}[1]{\mathcolor{hsblue4}{\textit{\ensuremath{#1}}}}
\newcommand{\HSCon}[1]{\mathcolor{hsblue3}{\mathit{\ensuremath{#1}}}}
\newcommand{\HSVar}[1]{\mathcolor{hsblue5}{\mathit{\ensuremath{#1}}}}

\newcommand{\HSComment}[1]{\mathcolor{hsgold2}{\textit{#1}}}

\newcommand{\HT}[1]{\ifdefined\HSCon\HSCon{#1}\else#1\fi}

\newcommand{\mathnocolor}[2]{#2}
\newcommand{\HSCustomNC}[2]{%
\mathcolor{#1}{\let\mathcolor\mathnocolor%
\ensuremath{#2}}}

\usepackage{tikz}
\usetikzlibrary{positioning}
\usetikzlibrary{calc}
\usetikzlibrary{plotmarks}
\usetikzlibrary{fit}
\usetikzlibrary{shapes}

\usepackage{cleveref}

\renewcommand{\!}{\negthinspace}

\definecolor{C1}{RGB}{0,153,204}
\definecolor{C2}{RGB}{89,0,179}

\newcounter{commentctr}
\renewcommand{\hscodestyle}{\footnotesize}

\makeatletter
  {\hsnewpar\abovedisplayskip
   \hscodestyle
   \let\hspre\(\let\hspost\)%
   \pboxed}
  {\endpboxed%
   \hsnewpar\belowdisplayskip
   \ignorespacesafterend}
  {\vspace*{-5.7em}\hscodestyle
   \let\hspre\(\let\hspost\)%
   \pboxed}
  {\endpboxed\ignorespacesafterend}
\makeatother
\sethscode{myownhscode}

\newenvironment{myhs*}[1][0.95\textwidth]{%
\begin{minipage}{#1}%
}{%
\end{minipage}%
}

\newcommand{\lhsinclude}[1]{}

\usepackage{subcaption}
\begin{document}
\title{Proof Engineering with Predicate Transformer Semantics}
\author{
  Christa Jenkins\inst{1,3}\orcidID{0000-0002-5434-5018} %
  \and Mark Moir\inst{2} %
  \and Harold Carr\inst{3} %
}
\authorrunning{C.\ Jenkins, M.\ Moir, H.\ Carr}
\institute{
  The University of Iowa, Iowa City, Iowa, USA %
      \email{christa-jenkins@uiowa.edu}   %
  \and Oracle Labs, New Zealand                %
  \and Oracle Labs, USA                        %
}
\maketitle              %
\begin{abstract}
  We present a lightweight, open source Agda framework for manually verifying effectful programs
  using predicate transformer semantics.
  We represent the abstract syntax trees (AST) of effectful programs with a
  generalized algebraic datatype (GADT) \ensuremath{\HSCon{AST}}, whose generality enables even
  complex operations to be primitive AST nodes.
  Users can then assign bespoke predicate transformers to such operations to aid
  the proof effort, for example by automatically decomposing proof obligations
  for branching code.
  Our framework codifies and generalizes a proof engineering methodology used by
  the authors to reason about a prototype implementation of \textsc{LibraBFT}, a
  Byzantine fault tolerant consensus protocol in which code executed by
  participants may have effects such as updating state and sending messages.
  Successful use of our framework in
  this context demonstrates its practical applicability.
\end{abstract}

\lhsinclude{abstract.lhs}
\section{Introduction}
\label{sec:introduction}

Interactive theorem provers (ITPs) based on dependent type theory provide a
flexible way to formally verify program properties, as they unify the
language of specification and computation into an expressive pure functional
programming language~\cite{SU06_Lectures-on-the-Curry-Howard-Isomorphism}.
By virtue of referential transparency, properties of pure functions (those
without side effects) can be proven with equational reasoning.
However, when the computation of interest is inherently effectful, other
techniques may be required.
For example, in the case of distributed systems, participants
perform state updates, emit messages, and invoke subroutines that may throw
exceptions; the network's ability to tolerate faults rests upon participants'
behaviors.

One approach to reasoning about effects, described by
Swierstra and Baanen~\cite{SB19_A-Predicate-Transformer-Semantics-for-Effects},
is to model the abstract syntax tree (AST) of the effectful program with a
datatype and assign to this type both an operational semantics and
\emph{predicate transformer} semantics (PTS).
PTS provides a structured way for reasoning about effectful code
\cite{SB19_A-Predicate-Transformer-Semantics-for-Effects,SWSCL13_Verifying-Higher-Order-Programs-Dijkstra-Monad},
reducing the goal of showing a given postcondition \(P\) holds of a program
\(m\) to proving the  weakest-precondition \(\mathit{wp}\ m\ P\) of \(P\)
w.r.t.\ that program.
As \(\mathit{wp}\ m\) maps \emph{arbitrary} postconditions to preconditions,
we may view it as giving a meaning (semantics) to \(m\) in terms of functions
from postconditions to preconditions (predicate transformers).

\paragraph*{Contributions}
In this paper, we describe a generic Agda framework for manually verifying
effectful programs using PTS.
This framework is designed to reduce the mental overhead for proof engineers
(hereafter, ``users'')
by tailoring the phrasing of intermediate proof obligations.
In particular, our framework allows directly assigning predicate transformers
not only to expected effectful operations, but also to monadic bind and pure
operations for branching.
Careful phrasing of proof obligations also facilitates a limited form of proof
synthesis when the goal type can be decomposed with a unique type-correct
constructor.

Our framework was developed as a part of our efforts to verify safety properties
of an implementation of \textsc{LibraBFT} in Agda.
We have previously reported on some aspects of that work~\cite{NASAFM-2022}.
Details of the work presented in this paper are available in the same open source
repository~\cite{librabft-agda}.
\textsc{LibraBFT} (a.k.a. \textsc{DiemBFT})~\cite{libra-2019-06-28} is a
real-world Byzantine-fault tolerant consensus protocol.
We believe the techniques we describe generalize to other domains and
ITPs, including those with greater levels of
automation~\cite{AHMMPPRS17_Dijkstra-Monads-for-Free}.

\section{Proof Engineering with Predicate Transformers}
\label{sec:proof-engineering-pts}

\begin{figure}[t]
  \centering
  \begin{hscode}\SaveRestoreHook
\column{B}{@{}>{\hspre}l<{\hspost}@{}}%
\column{3}{@{}>{\hspre}l<{\hspost}@{}}%
\column{15}{@{}>{\hspre}l<{\hspost}@{}}%
\column{22}{@{}>{\hspre}l<{\hspost}@{}}%
\column{34}{@{}>{\hspre}l<{\hspost}@{}}%
\column{E}{@{}>{\hspre}l<{\hspost}@{}}%
\>[B]{}\HSVar{prog}\;\mathbin{\HSCon{:}}\;\HSSpecial{(}\HSCon{St}\;\mathrel{\HSSym{\to}} \;\HSCon{Maybe}\;\HSCon{Wr}\HSSpecial{)}\;\mathrel{\HSSym{\to}} \;\HSCon{RWS}\;\HSCon{Unit}{}\<[E]%
\\
\>[B]{}\HSVar{prog}\;\HSVar{g}\;\mathrel{\HSSym{=}}\;\HSVar{pass}\;\HSVar{inner}\;\HSKeyword{where}{}\<[E]%
\\
\>[B]{}\hsindent{3}{}\<[3]%
\>[3]{}\HSVar{inner}\;\mathbin{\HSCon{:}}\;\HSCon{RWS}\;\HSSpecial{(}\HSCon{Unit}\;\mathbin{\HT{\times}}\;\HSSpecial{(}\HSCon{List}\;\HSCon{Wr}\;\mathrel{\HSSym{\to}} \;\HSCon{List}\;\HSCon{Wr}\HSSpecial{)}\HSSpecial{)}{}\<[E]%
\\
\>[B]{}\hsindent{3}{}\<[3]%
\>[3]{}\HSVar{inner}\;\mathrel{\HSSym{=}}\;\HSKeyword{do}\;{}\<[15]%
\>[15]{}\HSVar{m}\;\mathbin{\HSSym{\leftarrow}} \;\HSVar{gets}\;\HSVar{g}\;{}\<[E]%
\\
\>[15]{}\HSVar{maybe}\;{}\<[22]%
\>[22]{}\HSSpecial{(}\HSSym{\lambda} \;\HSVar{w}\;\mathrel{\HSSym{\to}} \;\HSKeyword{do}\;{}\<[34]%
\>[34]{}\HSVar{tell}\;\HSSym{[\mskip1.5mu} \;\HSVar{w}\;\HSSym{\mskip1.5mu]}\;{}\<[E]%
\\
\>[34]{}\HSVar{return}\;\HSSpecial{(}\HT{unit}\;\mathbin{\HSSym{,}}\;\HSSym{\lambda} \;\mathbin{\HSSym{\anonymous}} \;\mathrel{\HSSym{\to}} \;\HSVar{[]}\HSSpecial{)}\HSSpecial{)}\;{}\<[E]%
\\
\>[22]{}\HSSpecial{(}\HSVar{return}\;\HSSpecial{(}\HT{unit}\;\mathbin{\HSSym{,}}\;\HSSym{\lambda} \;\HSVar{x}\;\mathrel{\HSSym{\to}} \;\HSVar{x}\;\mathbin{\HSSym{\plus}} \;\HSVar{x}\HSSpecial{)}\HSSpecial{)}\;\HSVar{m}{}\<[E]%
\ColumnHook
\end{hscode}\resethooks
  \caption{An example effectful program}
  \label{fig:ex-prog1}
\end{figure}

As motivation, we consider the small effectful program \ensuremath{\HSVar{prog}} in
Figure~\ref{fig:ex-prog1}.
The type of \ensuremath{\HSVar{prog}\;\HSVar{g}} is \ensuremath{\HSCon{RWS}\;\HSCon{Unit}}\footnote{Many names in the
  repository---including \ensuremath{\HSCon{RWS}}---have \ensuremath{\HSCon{AST}} suffixes for disambiguation, which
  are usually omitted in this paper for brevity.}, which says that it produces
no interesting values (\ensuremath{\HSCon{Unit}} is the unitary type) and that it uses the
effects of the \emph{reader, writer, state} monad~\cite{RWS}: it may read and
write to state of type \ensuremath{\HSCon{St}}, emit messages of type \ensuremath{\HSCon{Wr}}, and read from an
environment of type \ensuremath{\HSCon{Ev}}.
To avoid confusion, we say that an effectful program of type \ensuremath{\HSCon{RWS}\;\HSCon{A}}
\emph{returns} or \emph{produces} a value of type \ensuremath{\HSCon{A}}, as opposed to
\emph{emitting} messages (always of type \ensuremath{\HSCon{List}\;\HSCon{Wr}}).
We assume the reader is familiar with Haskell-style \ensuremath{\HSKeyword{do}}-notation and the basics
of dependent type theory; we explain Agda-specific syntax and conventions as
they are introduced.

Given a function \ensuremath{\HSVar{g}\;\mathbin{\HSCon{:}}\;\HSCon{St}\;\mathrel{\HSSym{\to}} \;\HSCon{Maybe}\;\HSCon{Wr}} that may compute a message from a
state, \ensuremath{\HSVar{prog}\;\HSVar{g}} calls \ensuremath{\HSVar{pass}} on \ensuremath{\HSVar{inner}}; \ensuremath{\HSVar{pass}} has the effect of
applying the function of type
\ensuremath{\HSCon{List}\;\HSCon{Wr}\;\mathrel{\HSSym{\to}} \;\HSCon{List}\;\HSCon{Wr}} returned by \ensuremath{\HSVar{inner}} to the messages emitted
by \ensuremath{\HSVar{inner}}, the result of which is then emitted by \ensuremath{\HSVar{prog}\;\HSVar{g}}.
In \ensuremath{\HSVar{inner}}, we use \ensuremath{\HSVar{gets}} to apply \ensuremath{\HSVar{g}} to the current state, binding the result as
\ensuremath{\HSVar{m}\;\mathbin{\HSCon{:}}\;\HSCon{Maybe}\;\HSCon{Wr}}.
Using the \ensuremath{\HSVar{maybe}} operation to scrutinize \ensuremath{\HSVar{m}},
if a message item \ensuremath{\HSVar{w}\;\mathbin{\HSCon{:}}\;\HSCon{Wr}} is produced by \ensuremath{\HSVar{g}} (i.e., \ensuremath{\HSVar{m}\;\mathbin{\HT{\equiv}}\;\HT{just}\;\HSVar{w}}), then we emit it and
return a constant function that returns the empty list, in effect erasing what
was just emitted.
In case \ensuremath{\HSVar{m}\;\mathbin{\HT{\equiv}}\;\HT{nothing}}, \ensuremath{\HSVar{prog}\;\HSVar{g}} returns a function that concatenates a list of
messages with itself.

The operational semantics for \ensuremath{\HSCon{RWS}} programs is given by \ensuremath{\HSVar{runRWS}}, whose type is
shown below (the definition of \ensuremath{\HSVar{runRWS}} is in module \ensuremath{\HSCon{Dijkstra.AST.RWS}}; module
\ensuremath{\HSCon{X.Y.Z}} lives in \texttt{src/X/Y/Z.agda} in~\cite{librabft-agda}).

\begin{hscode}\SaveRestoreHook
\column{B}{@{}>{\hspre}l<{\hspost}@{}}%
\column{11}{@{}>{\hspre}l<{\hspost}@{}}%
\column{E}{@{}>{\hspre}l<{\hspost}@{}}%
\>[B]{}\HSCon{Input}\;{}\<[11]%
\>[11]{}\mathrel{\HSSym{=}}\;\HSCon{Ev}\;\mathbin{\HT{\times}}\;\HSCon{St}{}\<[E]%
\\
\>[B]{}\HSCon{Output}\;\HSCon{A}\;{}\<[11]%
\>[11]{}\mathrel{\HSSym{=}}\;\HSCon{A}\;\mathbin{\HT{\times}}\;\HSCon{St}\;\mathbin{\HT{\times}}\;\HSCon{List}\;\HSCon{Wr}{}\<[E]%
\\[\blanklineskip]%
\>[B]{}\HSVar{runRWS}\;\mathbin{\HSCon{:}}\;\HSSym{\forall}\;\HSSpecial{\HSSym{\{\mskip1.5mu} }\HSCon{A}\HSSpecial{\HSSym{\mskip1.5mu\}}}\;\mathrel{\HSSym{\to}} \;\HSCon{RWS}\;\HSCon{A}\;\mathrel{\HSSym{\to}} \;\HSCon{Input}\;\mathrel{\HSSym{\to}} \;\HSCon{Output}\;\HSCon{A}{}\<[E]%
\ColumnHook
\end{hscode}\resethooks
To run a program of type \ensuremath{\HSCon{RWS}\;\HSCon{A}}, \ensuremath{\HSVar{runRWS}} requires as input an environment
value and prestate, and the result of executing such a program is a triple
consisting of a value of type \ensuremath{\HSCon{A}}, a poststate, and a list of emitted messages.

\subsubsection*{Direct Approach to Proof Engineering}
\label{sec:prog-direct}

Suppose we are tasked with verifying that \ensuremath{\HSVar{prog}}
satisfies postcondition \ensuremath{\HSCon{ProgPost}} below, which expresses the property that the
prestate and poststate are equal and that there are no outputs.
\begin{hscode}\SaveRestoreHook
\column{B}{@{}>{\hspre}l<{\hspost}@{}}%
\column{E}{@{}>{\hspre}l<{\hspost}@{}}%
\>[B]{}\HSCon{ProgPost}\;\mathbin{\HSCon{:}}\;\HSCon{Input}\;\mathrel{\HSSym{\to}} \;\HSCon{Output}\;\HSCon{Unit}\;\mathrel{\HSSym{\to}} \;\HSCon{Set}{}\<[E]%
\\
\>[B]{}\HSCon{ProgPost}\;\HSSpecial{(}\HSVar{\mathit{e}_{1}}\;\mathbin{\HSSym{,}}\;\HSVar{\mathit{s}_{1}}\HSSpecial{)}\;\HSSpecial{(}\HSVar{\mathit{u}_{1}}\;\mathbin{\HSSym{,}}\;\HSVar{\mathit{s}_{2}}\;\mathbin{\HSSym{,}}\;\HSVar{o}\HSSpecial{)}\;\mathrel{\HSSym{=}}\;\HSVar{\mathit{s}_{1}}\;\mathbin{\HT{\equiv}}\;\HSVar{\mathit{s}_{2}}\;\mathbin{\HT{\times}}\;\HSVar{0}\;\mathbin{\HT{\equiv}}\;\HSVar{length}\;\HSVar{o}{}\<[E]%
\ColumnHook
\end{hscode}\resethooks
(\ensuremath{\HSCon{Set}} is a classifier for types in Agda; the type of \ensuremath{\HSCon{ProgPost}}
says that it is a relation between inputs to, and outputs of,
programs of type \ensuremath{\HSCon{RWS}\;\HSCon{Unit}}.)

Attempting the proof directly, we begin with:
\begin{hscode}\SaveRestoreHook
\column{B}{@{}>{\hspre}l<{\hspost}@{}}%
\column{E}{@{}>{\hspre}l<{\hspost}@{}}%
\>[B]{}\HSVar{progPost}\;\mathbin{\HSCon{:}}\;\HSSym{\forall}\;\HSVar{g}\;\HSVar{i}\;\mathrel{\HSSym{\to}} \;\HSCon{ProgPost}\;\HSVar{i}\;\HSSpecial{(}\HSVar{runRWS}\;\HSSpecial{(}\HSVar{prog}\;\HSVar{g}\HSSpecial{)}\;\HSVar{i}\HSSpecial{)}{}\<[E]%
\\
\>[B]{}\HSVar{progPost}\;\HSVar{g}\;\HSSpecial{(}\HSVar{e}\;\mathbin{\HSSym{,}}\;\HSVar{s}\HSSpecial{)}\;\mathrel{\HSSym{=}}\;\HSVar{?}{}\<[E]%
\ColumnHook
\end{hscode}\resethooks
where the \ensuremath{\HSVar{?}} marks a hole in the proof.
When Agda unfolds the proof obligation at the hole,
the result is unwieldy!
To give an idea, here is (a cleaned-up version of) the obligation for
just the second conjunct of the postcondition \ensuremath{\HSCon{ProgPost}}:
\begin{hscode}\SaveRestoreHook
\column{B}{@{}>{\hspre}l<{\hspost}@{}}%
\column{3}{@{}>{\hspre}l<{\hspost}@{}}%
\column{9}{@{}>{\hspre}l<{\hspost}@{}}%
\column{12}{@{}>{\hspre}l<{\hspost}@{}}%
\column{14}{@{}>{\hspre}l<{\hspost}@{}}%
\column{16}{@{}>{\hspre}l<{\hspost}@{}}%
\column{E}{@{}>{\hspre}l<{\hspost}@{}}%
\>[3]{}\HSVar{0}\;\mathbin{\HT{\equiv}}\;{}\<[9]%
\>[9]{}\HSVar{length}\;\HSSpecial{(}\HSVar{snd}\;\HSSpecial{(}\HSVar{fst}\;\HSSpecial{(}\HSVar{runRWS}{}\<[E]%
\\
\>[9]{}\hsindent{3}{}\<[12]%
\>[12]{}\HSSpecial{(}\HSVar{maybe}\;\HSSpecial{(}\HSVar{return}\;\HSSpecial{(}\HT{unit}\;\mathbin{\HSSym{,}}\;\HSSym{\lambda} \;\HSVar{x}\;\mathrel{\HSSym{\to}} \;\HSVar{x}\;\mathbin{\HSSym{\plus}} \;\HSVar{x}\HSSpecial{)}\HSSpecial{)}\;{}\<[E]%
\\
\>[12]{}\hsindent{2}{}\<[14]%
\>[14]{}\HSSpecial{(}\HSSym{\lambda} \;\HSVar{w}\;\mathrel{\HSSym{\to}} \;\HSVar{tell}\;\HSSym{[\mskip1.5mu} \;\HSVar{w}\;\HSSym{\mskip1.5mu]}\;\mathbin{\HSSym{\sequ}} \;\HSVar{return}\;\HSSpecial{(}\HT{unit}\;\mathbin{\HSSym{,}}\;\HSSym{\lambda} \;\mathbin{\HSSym{\anonymous}} \;\mathrel{\HSSym{\to}} \;\HSVar{[]}\HSSpecial{)}\HSSpecial{)}\;\HSSpecial{(}\HSVar{g}\;\HSVar{s}\HSSpecial{)}\HSSpecial{)}\;\HSSpecial{(}\HSVar{e}\;\mathbin{\HSSym{,}}\;\HSVar{s}\HSSpecial{)}\HSSpecial{)}\HSSpecial{)}{}\<[E]%
\\
\>[9]{}\hsindent{3}{}\<[12]%
\>[12]{}\HSSpecial{(}\HSVar{snd}\;\HSSpecial{(}\HSVar{snd}\;\HSSpecial{(}\HSVar{runRWS}{}\<[E]%
\\
\>[12]{}\hsindent{2}{}\<[14]%
\>[14]{}\HSSpecial{(}\HSVar{maybe}\;\HSSpecial{(}\HSVar{return}\;\HSSpecial{(}\HT{unit}\;\mathbin{\HSSym{,}}\;\HSSym{\lambda} \;\HSVar{x}\;\mathrel{\HSSym{\to}} \;\HSVar{x}\;\mathbin{\HSSym{\plus}} \;\HSVar{x}\HSSpecial{)}\HSSpecial{)}\;{}\<[E]%
\\
\>[14]{}\hsindent{2}{}\<[16]%
\>[16]{}\HSSpecial{(}\HSSym{\lambda} \;\HSVar{w}\;\mathrel{\HSSym{\to}} \;\HSVar{tell}\;\HSSym{[\mskip1.5mu} \;\HSVar{w}\;\HSSym{\mskip1.5mu]}\;\mathbin{\HSSym{\sequ}} \;\HSVar{return}\;\HSSpecial{(}\HT{unit}\;\mathbin{\HSSym{,}}\;\HSSym{\lambda} \;\mathbin{\HSSym{\anonymous}} \;\mathrel{\HSSym{\to}} \;\HSVar{[]}\HSSpecial{)}\HSSpecial{)}\;\HSSpecial{(}\HSVar{g}\;\HSVar{s}\HSSpecial{)}\HSSpecial{)}\;\HSSpecial{(}\HSVar{e}\;\mathbin{\HSSym{,}}\;\HSVar{s}\HSSpecial{)}\HSSpecial{)}\HSSpecial{)}\HSSpecial{)}\HSSpecial{)}{}\<[E]%
\ColumnHook
\end{hscode}\resethooks
\noindent
(The proofs in this section are in the \ensuremath{\HSCon{Dijkstra.AST.Examples.PaperIntro}} module with more detail,
including the entire goal for the hole above in gory detail that a user attempting a direct
proof encounters.)

This obligation has parts of the program text from \ensuremath{\HSVar{inner}} repeated
\emph{twice,} once for the function it returns and once for the messages it
emits.
The issue is that the execution of \ensuremath{\HSVar{prog}} is stuck on the branching operation
\ensuremath{\HSVar{maybe}} whose scrutinee \ensuremath{\HSVar{g}\;\HSVar{s}} is not in weak head normal form (that is, it is
not of the form \ensuremath{\HT{nothing}} or \ensuremath{\HT{just}\;\HSVar{x}}).
Because our proof obligation refers directly to the evaluation of \ensuremath{\HSVar{prog}}
twice, the continuation in \ensuremath{\HSVar{inner}} is also referenced twice.

One way to proceed in this case is
to use the \ensuremath{\HSKeyword{with}} keyword to abstract over the expression blocking progress:
\begin{hscode}\SaveRestoreHook
\column{B}{@{}>{\hspre}l<{\hspost}@{}}%
\column{6}{@{}>{\hspre}l<{\hspost}@{}}%
\column{17}{@{}>{\hspre}l<{\hspost}@{}}%
\column{E}{@{}>{\hspre}l<{\hspost}@{}}%
\>[B]{}\HSVar{progPost}\;\HSVar{g}\;\HSSpecial{(}\HSVar{e}\;\mathbin{\HSSym{,}}\;\HSVar{s}\HSSpecial{)}\;\HSKeyword{with}\;\HSVar{g}\;\HSVar{s}{}\<[E]%
\\
\>[B]{}\HSVar{...}\;{}\<[6]%
\>[6]{}\mathbin{\HSSym{\mid}} \;\HT{nothing}\;{}\<[17]%
\>[17]{}\mathrel{\HSSym{=}}\;\HSVar{?}{}\<[E]%
\\
\>[B]{}\HSVar{...}\;{}\<[6]%
\>[6]{}\mathbin{\HSSym{\mid}} \;\HT{just}\;\HSVar{w}\;{}\<[17]%
\>[17]{}\mathrel{\HSSym{=}}\;\HSVar{?}{}\<[E]%
\ColumnHook
\end{hscode}\resethooks
with the result that the new subgoals this generates (\ensuremath{\HSVar{s}\;\mathbin{\HT{\equiv}}\;\HSVar{s}\;\mathbin{\HT{\times}}\;\HSVar{0}\;\mathbin{\HT{\equiv}}\;\HSVar{0}} in
both cases) are much easier to understand than the original goal.

Though \ensuremath{\HSVar{prog}} is a small example, and the property we wish to verify is
relatively simple, the preceding discussion illustrates difficulties that are
greatly magnified as the complexity of the task increases.
First, proof obligations for ``stuck'' code can explode in size, especially when
the code branches, making it difficult to read the proof state and identify
\emph{why} execution is stuck.
Second, once identified, using \ensuremath{\HSKeyword{with}} to inspect the cases of the scrutinee of
branching code means recapitulating the effectful operations performed up to that point.
In the example considered above, there were no updates to the state \ensuremath{\HSVar{s}} before
the \ensuremath{\HSVar{gets}} operation was used, but had there been for example a preceding \ensuremath{\HSVar{puts}\;\HSVar{p}} (for some \ensuremath{\HSVar{p}\;\mathbin{\HSCon{:}}\;\HSCon{St}\;\mathrel{\HSSym{\to}} \;\HSCon{St}}), the user would instead have to write
\ensuremath{\HSKeyword{with}\;\HSVar{g}\;\HSSpecial{(}\HSVar{p}\;\HSVar{s}\HSSpecial{)}}.

\subsubsection*{Simplifying Proof Obligations with PTS}
Our Agda framework is designed to address these issues.
Following the general approach of Swierstra and
Baanen~\cite{SB19_A-Predicate-Transformer-Semantics-for-Effects}, our framework
uses a datatype for the ASTs of effectful computations (\ensuremath{\HSCon{AST}},
Figure~\ref{fig:ast-gadt}); \ensuremath{\HSCon{RWS}} is an instance of this datatype.
To prove that \ensuremath{\HSCon{ProgPost}} holds of \ensuremath{\HSVar{prog}}, we first prove a \emph{precondition}
obtained by the function
\begin{hscode}\SaveRestoreHook
\column{B}{@{}>{\hspre}l<{\hspost}@{}}%
\column{E}{@{}>{\hspre}l<{\hspost}@{}}%
\>[B]{}\HSVar{predTrans}\;\mathbin{\HSCon{:}}\;\HSSym{\forall}\;\HSSpecial{\HSSym{\{\mskip1.5mu} }\HSCon{A}\HSSpecial{\HSSym{\mskip1.5mu\}}}\;\mathrel{\HSSym{\to}} \;\HSCon{RWS}\;\HSCon{A}\;\mathrel{\HSSym{\to}} \;\HSSpecial{(}\HSCon{Output}\;\HSCon{A}\;\mathrel{\HSSym{\to}} \;\HSCon{Set}\HSSpecial{)}\;\mathrel{\HSSym{\to}} \;\HSCon{Input}\;\mathrel{\HSSym{\to}} \;\HSCon{Set}{}\<[E]%
\ColumnHook
\end{hscode}\resethooks
which assigns to each \ensuremath{\HSCon{RWS}} program (by induction over its AST) a function that
maps postconditions to preconditions.
Put another way, it gives each \ensuremath{\HSCon{RWS}} program a \emph{semantics} as a
\emph{predicate transformer,} as it transforms predicates over \ensuremath{\HSCon{Output}} types to
predicates over the \ensuremath{\HSCon{Input}} type.

We begin our proof that precondition for \ensuremath{\HSCon{ProgPost}} holds with:
\begin{hscode}\SaveRestoreHook
\column{B}{@{}>{\hspre}l<{\hspost}@{}}%
\column{E}{@{}>{\hspre}l<{\hspost}@{}}%
\>[B]{}\HSVar{progPostWP}\;\mathbin{\HSCon{:}}\;\HSSym{\forall}\;\HSVar{g}\;\HSVar{i}\;\mathrel{\HSSym{\to}} \;\HSVar{predTrans}\;\HSSpecial{(}\HSVar{prog}\;\HSVar{g}\HSSpecial{)}\;\HSSpecial{(}\HSCon{ProgPost}\;\HSVar{i}\HSSpecial{)}\;\HSVar{i}{}\<[E]%
\\
\>[B]{}\HSVar{progPostWP}\;\HSVar{g}\;\HSSpecial{(}\HSVar{e}\;\mathbin{\HSSym{,}}\;\HSVar{s}\HSSpecial{)}\;\mathrel{\HSSym{=}}\;\HSVar{?}{}\<[E]%
\ColumnHook
\end{hscode}\resethooks
\noindent where the proof obligation at the hole is now much more understandable:
\begin{hscode}\SaveRestoreHook
\column{B}{@{}>{\hspre}l<{\hspost}@{}}%
\column{5}{@{}>{\hspre}l<{\hspost}@{}}%
\column{12}{@{}>{\hspre}l<{\hspost}@{}}%
\column{16}{@{}>{\hspre}l<{\hspost}@{}}%
\column{E}{@{}>{\hspre}l<{\hspost}@{}}%
\>[B]{}\HSSpecial{(}\HSVar{r}\;{}\<[5]%
\>[5]{}\mathbin{\HSCon{:}}\;\HSCon{Maybe}\;\HSCon{Wr}\HSSpecial{)}\;\mathrel{\HSSym{\to}} \;\HSVar{r}\;\mathbin{\HT{\equiv}}\;\HSVar{g}\;\HSVar{s}\;\mathrel{\HSSym{\to}} \;{}\<[E]%
\\
\>[5]{}\hsindent{7}{}\<[12]%
\>[12]{}\HSSpecial{(}\HSSpecial{(}\HSVar{j}\;\mathbin{\HSCon{:}}\;\HSCon{Wr}\HSSpecial{)}\;\mathrel{\HSSym{\to}} \;\HSVar{r}\;\mathbin{\HT{\equiv}}\;\HT{just}\;\HSVar{j}\;\mathrel{\HSSym{\to}} \;\HSSpecial{(}\HSVar{\mathit{r}_{1}}\;\mathbin{\HSCon{:}}\;\HSCon{Unit}\HSSpecial{)}\;\mathrel{\HSSym{\to}} \;\HSVar{\mathit{r}_{1}}\;\mathbin{\HT{\equiv}}\;\HT{unit}\;\mathrel{\HSSym{\to}} \;{}\<[E]%
\\
\>[12]{}\hsindent{4}{}\<[16]%
\>[16]{}\HSSpecial{(}\HSVar{o'}\;\mathbin{\HSCon{:}}\;\HSCon{List}\;\HSCon{Wr}\HSSpecial{)}\;\mathrel{\HSSym{\to}} \;\HSVar{o'}\;\mathbin{\HT{\equiv}}\;\HSVar{[]}\;\mathrel{\HSSym{\to}} \;\HSSpecial{(}\HSVar{s}\;\mathbin{\HT{\equiv}}\;\HSVar{s}\HSSpecial{)}\;\mathbin{\HT{\times}}\;\HSSpecial{(}\HSVar{0}\;\mathbin{\HT{\equiv}}\;\HSVar{length}\;\HSVar{o'}\HSSpecial{)}\HSSpecial{)}\;{}\<[E]%
\\
\>[5]{}\mathbin{\HT{\times}}\;{}\<[12]%
\>[12]{}\HSSpecial{(}\HSVar{r}\;\mathbin{\HT{\equiv}}\;\HT{nothing}\;\mathrel{\HSSym{\to}} \;\HSSpecial{(}\HSVar{o'}\;\mathbin{\HSCon{:}}\;\HSCon{List}\;\HSCon{Wr}\HSSpecial{)}\;\mathrel{\HSSym{\to}} \;\HSVar{o'}\;\mathbin{\HT{\equiv}}\;\HSVar{[]}\;\mathrel{\HSSym{\to}} \;{}\<[E]%
\\
\>[12]{}\hsindent{4}{}\<[16]%
\>[16]{}\HSSpecial{(}\HSVar{s}\;\mathbin{\HT{\equiv}}\;\HSVar{s}\HSSpecial{)}\;\mathbin{\HT{\times}}\;\HSSpecial{(}\HSVar{0}\;\mathbin{\HT{\equiv}}\;\HSVar{length}\;\HSVar{o'}\HSSpecial{)}\HSSpecial{)}{}\<[E]%
\ColumnHook
\end{hscode}\resethooks

The user's next few steps are entirely type
directed: introduce the premises of the implication to the context, then (because
this leaves us with a product) decompose the proof obligation into two
subobligations (the detailed commentary in \ensuremath{\HSCon{Dijkstra.AST.Examples.PaperIntro}}
shows how to develop the proof largely automatcally using Emacs's Agda mode).
This yields:
\begin{hscode}\SaveRestoreHook
\column{B}{@{}>{\hspre}l<{\hspost}@{}}%
\column{E}{@{}>{\hspre}l<{\hspost}@{}}%
\>[B]{}\HT{proj_{1}}\;\HSSpecial{(}\HSVar{progPostWP}\;\HSVar{g}\;\HSSpecial{(}\HSVar{e}\;\mathbin{\HSSym{,}}\;\HSVar{s}\HSSpecial{)}\;\HSVar{m}\;\HSVar{mId}\HSSpecial{)}\;\mathrel{\HSSym{=}}\;\HSVar{?}{}\<[E]%
\\
\>[B]{}\HT{proj_{2}}\;\HSSpecial{(}\HSVar{progPostWP}\;\HSVar{g}\;\HSSpecial{(}\HSVar{e}\;\mathbin{\HSSym{,}}\;\HSVar{s}\HSSpecial{)}\;\HSVar{m}\;\HSVar{mId}\HSSpecial{)}\;\mathrel{\HSSym{=}}\;\HSVar{?}{}\<[E]%
\ColumnHook
\end{hscode}\resethooks

At this point, we note two things.
First, the framework has enabled us to alias \ensuremath{\HSVar{g}\;\HSVar{s}} as \ensuremath{\HSVar{m}} (we 
choose the name \ensuremath{\HSVar{m}}, rather than \ensuremath{\HSVar{r}} as shown in the proof obligation, to
better match the local name in \ensuremath{\HSVar{prog}}), and the remaining proof obligations
\emph{only} mention \ensuremath{\HSVar{m}}.
Term \ensuremath{\HSVar{mId}\;\mathbin{\HSCon{:}}\;\HSVar{m}\;\mathbin{\HT{\equiv}}\;\HSVar{g}\;\HSVar{s}} enables us to undo this aliasing as needed, as
dependent pattern matching on \ensuremath{\HSVar{mId}} replaces \ensuremath{\HSVar{m}} with \ensuremath{\HSVar{g}\;\HSVar{s}} in the proof state.
Recall that, in the direct proof above, \ensuremath{\HSVar{g}\;\HSVar{s}} was repeated in the proof
obligation. 
If we had a more complex expression as the scrutinee, it too would be repeated
in the direct proof---but in a proof using our framework, \emph{only the alias}
is repeated!
Second, and relatedly, in the direct approach, the user must
explicitly invoke \ensuremath{\HSKeyword{with}} on the scrutinee \ensuremath{\HSVar{g}\;\HSVar{s}} in order to generate two
subobligations (one each for \ensuremath{\HT{nothing}} and \ensuremath{\HT{just}}).
With our framework, these two cases are \emph{already} present in the proof
obligation, in the form of a product of obligations in which, while proving each
component, the user assumes that the alias \ensuremath{\HSVar{m}} is either \ensuremath{\HT{nothing}} or \ensuremath{\HT{just}\;\HSVar{w}}
for some \ensuremath{\HSVar{w}\;\mathbin{\HSCon{:}}\;\HSCon{Wr}}.

The rest of the proof is similarly type-directed by the framework and is fairly
straightforward; see \ensuremath{\HSCon{Dijkstra.AST.Examples.PaperIntro}} for the details.
Finally, having proved \ensuremath{\HSVar{progPostWP}}, we can
obtain a proof of the desired postcondition by using \ensuremath{\HSVar{sufficient}}, which is a proof
that the preconditions computed by \ensuremath{\HSVar{predTrans}} are \emph{sufficient} for proving the
given postcondition for the given program.
\begin{hscode}\SaveRestoreHook
\column{B}{@{}>{\hspre}l<{\hspost}@{}}%
\column{E}{@{}>{\hspre}l<{\hspost}@{}}%
\>[B]{}\HSVar{progPost}\;\mathbin{\HSCon{:}}\;\HSSym{\forall}\;\HSVar{g}\;\HSVar{i}\;\mathrel{\HSSym{\to}} \;\HSCon{ProgPost}\;\HSVar{i}\;\HSSpecial{(}\HSVar{runRWS}\;\HSSpecial{(}\HSVar{prog}\;\HSVar{g}\HSSpecial{)}\;\HSVar{i}\HSSpecial{)}{}\<[E]%
\\
\>[B]{}\HSVar{progPost}\;\HSVar{g}\;\HSVar{i}\;\mathrel{\HSSym{=}}\;\HSVar{sufficient}\;\HSSpecial{(}\HSVar{prog}\;\HSVar{g}\HSSpecial{)}\;\HSSpecial{(}\HSCon{ProgPost}\;\HSVar{i}\HSSpecial{)}\;\HSVar{i}\;\HSSpecial{(}\HSVar{progPostWP}\;\HSVar{g}\;\HSVar{i}\HSSpecial{)}{}\<[E]%
\ColumnHook
\end{hscode}\resethooks

\section{Framework for PTS}
\label{sec:pts-rws}

In this section, we describe our generic framework for modeling effectful
computations and reasoning about them with predicate transformer semantics.
We define a generalized algebraic datatype (GADT) for the ASTs of
effectful programs, parameterized by a collection of operations supplied by the
user.
To set up the framework for a particular set of effects, the user provides an
operational and predicate transformer semantics (PTS) for the operations, then
proves that these two semantics agree; this enables
verifying a postcondition by proving the sufficient
precondition generated by the PTS.
It is at the point of specifying the predicate transformer semantics that one
may tailor the computed proof obligations to suit one's needs, for example,
introducing aliasing or reducing the goal to some set of subgoals;
proof obligations can be rephrased in any convenient way,
provided the two semantics can be shown to agree.

\subsection{\ensuremath{\HSCon{AST}}}
\label{sec:ast}

\begin{figure}[t]
  \begin{hscode}\SaveRestoreHook
\column{B}{@{}>{\hspre}l<{\hspost}@{}}%
\column{3}{@{}>{\hspre}l<{\hspost}@{}}%
\column{5}{@{}>{\hspre}l<{\hspost}@{}}%
\column{13}{@{}>{\hspre}l<{\hspost}@{}}%
\column{14}{@{}>{\hspre}l<{\hspost}@{}}%
\column{16}{@{}>{\hspre}l<{\hspost}@{}}%
\column{61}{@{}>{\hspre}l<{\hspost}@{}}%
\column{E}{@{}>{\hspre}l<{\hspost}@{}}%
\>[B]{}\HSKeyword{record}\;\HSCon{ASTOps}\;\mathbin{\HSCon{:}}\;\HSCon{Set}\;\HSKeyword{where}{}\<[E]%
\\
\>[B]{}\hsindent{3}{}\<[3]%
\>[3]{}\HSKeyword{field}{}\<[E]%
\\
\>[3]{}\hsindent{2}{}\<[5]%
\>[5]{}\HSCon{Cmd}\;{}\<[13]%
\>[13]{}\mathbin{\HSCon{:}}\;\HSSpecial{(}\HSCon{A}\;\mathbin{\HSCon{:}}\;\HSCon{Set}\HSSpecial{)}\;\mathrel{\HSSym{\to}} \;\HSCon{Set}{}\<[E]%
\\
\>[3]{}\hsindent{2}{}\<[5]%
\>[5]{}\HSCon{SubArg}\;{}\<[13]%
\>[13]{}\mathbin{\HSCon{:}}\;\HSSpecial{\HSSym{\{\mskip1.5mu} }\HSCon{A}\;\mathbin{\HSCon{:}}\;\HSCon{Set}\HSSpecial{\HSSym{\mskip1.5mu\}}}\;\HSSpecial{(}\HSVar{c}\;\mathbin{\HSCon{:}}\;\HSCon{Cmd}\;\HSCon{A}\HSSpecial{)}\;\mathrel{\HSSym{\to}} \;\HSCon{Set}{}\<[E]%
\\
\>[3]{}\hsindent{2}{}\<[5]%
\>[5]{}\HSCon{SubRet}\;{}\<[13]%
\>[13]{}\mathbin{\HSCon{:}}\;\HSSpecial{\HSSym{\{\mskip1.5mu} }\HSCon{A}\;\mathbin{\HSCon{:}}\;\HSCon{Set}\HSSpecial{\HSSym{\mskip1.5mu\}}}\;\HSSpecial{\HSSym{\{\mskip1.5mu} }\HSVar{c}\;\mathbin{\HSCon{:}}\;\HSCon{Cmd}\;\HSCon{A}\HSSpecial{\HSSym{\mskip1.5mu\}}}\;\HSSpecial{(}\HSVar{r}\;\mathbin{\HSCon{:}}\;\HSCon{SubArg}\;\HSVar{c}\HSSpecial{)}\;\mathrel{\HSSym{\to}} \;\HSCon{Set}{}\<[E]%
\\
\>[B]{}\HSKeyword{open}\;\HSCon{ASTOps}{}\<[E]%
\\[\blanklineskip]%
\>[B]{}\HSKeyword{data}\;\HSCon{AST}\;\HSSpecial{(}\HSCon{OP}\;\mathbin{\HSCon{:}}\;\HSCon{ASTOps}\HSSpecial{)}\;\mathbin{\HSCon{:}}\;\HSCon{Set}\;\mathrel{\HSSym{\to}} \;\HSCon{Set}\;\HSKeyword{where}{}\<[E]%
\\
\>[B]{}\hsindent{3}{}\<[3]%
\>[3]{}\HSCon{ASTreturn}\;{}\<[14]%
\>[14]{}\mathbin{\HSCon{:}}\;\HSSym{\forall}\;\HSSpecial{\HSSym{\{\mskip1.5mu} }\HSCon{A}\HSSpecial{\HSSym{\mskip1.5mu\}}}\;\mathrel{\HSSym{\to}} \;\HSCon{A}\;{}\<[61]%
\>[61]{}\mathrel{\HSSym{\to}} \;\HSCon{AST}\;\HSCon{OP}\;\HSCon{A}{}\<[E]%
\\
\>[B]{}\hsindent{3}{}\<[3]%
\>[3]{}\HSCon{ASTbind}\;{}\<[14]%
\>[14]{}\mathbin{\HSCon{:}}\;\HSSym{\forall}\;\HSSpecial{\HSSym{\{\mskip1.5mu} }\HSCon{A}\;\HSCon{B}\HSSpecial{\HSSym{\mskip1.5mu\}}}\;\mathrel{\HSSym{\to}} \;\HSCon{AST}\;\HSCon{OP}\;\HSCon{A}\;\mathrel{\HSSym{\to}} \;\HSSpecial{(}\HSCon{A}\;\mathrel{\HSSym{\to}} \;\HSCon{AST}\;\HSCon{OP}\;\HSCon{B}\HSSpecial{)}\;{}\<[61]%
\>[61]{}\mathrel{\HSSym{\to}} \;\HSCon{AST}\;\HSCon{OP}\;\HSCon{B}{}\<[E]%
\\
\>[B]{}\hsindent{3}{}\<[3]%
\>[3]{}\HSCon{ASTop}\;{}\<[14]%
\>[14]{}\mathbin{\HSCon{:}}\;\HSSym{\forall}\;\HSSpecial{\HSSym{\{\mskip1.5mu} }\HSCon{A}\HSSpecial{\HSSym{\mskip1.5mu\}}}\;\mathrel{\HSSym{\to}} \;\HSSpecial{(}\HSVar{c}\;\mathbin{\HSCon{:}}\;\HSCon{Cmd}\;\HSCon{OP}\;\HSCon{A}\HSSpecial{)}{}\<[E]%
\\
\>[14]{}\hsindent{2}{}\<[16]%
\>[16]{}\mathrel{\HSSym{\to}} \;\HSSpecial{(}\HSVar{f}\;\mathbin{\HSCon{:}}\;\HSSpecial{(}\HSVar{r}\;\mathbin{\HSCon{:}}\;\HSCon{SubArg}\;\HSCon{OP}\;\HSVar{c}\HSSpecial{)}\;\mathrel{\HSSym{\to}} \;\HSCon{AST}\;\HSCon{OP}\;\HSSpecial{(}\HSCon{SubRet}\;\HSCon{OP}\;\HSVar{r}\HSSpecial{)}\HSSpecial{)}\;{}\<[E]%
\\
\>[16]{}\hsindent{45}{}\<[61]%
\>[61]{}\mathrel{\HSSym{\to}} \;\HSCon{AST}\;\HSCon{OP}\;\HSCon{A}{}\<[E]%
\ColumnHook
\end{hscode}\resethooks
  \caption{Datatype for effectful code}
  \label{fig:ast-gadt}
\end{figure}

Figure~\ref{fig:ast-gadt} shows the definition of \ensuremath{\HSCon{AST}}, the GADT for effectful
program ASTs (in Agda, \ensuremath{\HSCon{Set}} is a classifier for types, and the sort \ensuremath{\HSCon{Set}\;\mathrel{\HSSym{\to}} \;\HSCon{Set}} is for type constructors; to improve readability, our Agda code listings
omit universe levels~\cite{Agda22_Agda-Docs-Universe}).
These definitions are in the \ensuremath{\HSCon{Dijkstra.AST.Core}} module.
We make the monadic \emph{unit} operation (\ensuremath{\HSCon{ASTreturn}}) a constructor and, deviating from
the recipe of Hancock and Setzer~\cite{HS00_Interactive-Programs-in-DTT}, we also
make the \emph{bind} operation (\ensuremath{\HSCon{ASTbind}}) a constructor.
In Section~\ref{sec:ast-pts}, we will see that, by making the bind operator a
constructor, we avoid the need for an additional lemma to assign a predicate
transformer to composite computations (such as needed by Swierstra and
Baanen~\cite{SB19_A-Predicate-Transformer-Semantics-for-Effects}, \S 4).
The last constructor, \ensuremath{\HSCon{ASTop}}, enables us to describe effectful operations.

\ensuremath{\HSCon{AST}} is parameterized by a value of type \ensuremath{\HSCon{ASTOps}}, which comprises three
fields that describe the syntax of effectful operations.
First, \ensuremath{\HSCon{Cmd}} is the family of types for commands, indexed by a type \ensuremath{\HSCon{A}} for the
result value of the command.
Next, \ensuremath{\HSCon{SubArg}} is the family of argument types for the subcomputations (if any)
of a given command.
Finally, \ensuremath{\HSCon{SubRet}} is the family of types for values returned by subcomputations
(these need not be the same as the given type \ensuremath{\HSCon{A}} for the result type of the
whole command).
Curly braces around bound type variables indicate that we wish Agda to
infer instantiations of that type argument.

The \ensuremath{\HSCon{ASTop}} constructor of \ensuremath{\HSCon{AST}} takes as arguments a
command \ensuremath{\HSVar{c}} and a \ensuremath{\HSSpecial{(}\HSVar{r}\;\mathbin{\HSCon{:}}\;\HSCon{SubArg}\;\HSCon{OP}\;\HSVar{c}\HSSpecial{)}}-indexed family \ensuremath{\HSVar{f}} of subcomputations of the
command, each producing values of type \ensuremath{\HSCon{SubRet}\;\HSCon{OP}\;\HSVar{r}}, where the type of values
produced by the whole operation is \ensuremath{\HSCon{A}}.
As \ensuremath{\HSCon{AST}} has an explicit sequencing constructor (\ensuremath{\HSCon{ASTbind}}), we understand \ensuremath{\HSVar{f}}
not as a continuation for the next computation, but a family of computations
subordinate to the command itself (see Example~\ref{ex:ast-rws-op}).
These changes give us the flexibility to encode complex operations as primitive
nodes of the AST, enabling us to assign bespoke predicate transformers to them
when we reason about the behaviors of the programs in which they occur.

\begin{figure}[t]
  \begin{subfigure}{0.48\linewidth}
    \begin{hscode}\SaveRestoreHook
\column{B}{@{}>{\hspre}l<{\hspost}@{}}%
\column{3}{@{}>{\hspre}l<{\hspost}@{}}%
\column{5}{@{}>{\hspre}l<{\hspost}@{}}%
\column{12}{@{}>{\hspre}l<{\hspost}@{}}%
\column{E}{@{}>{\hspre}l<{\hspost}@{}}%
\>[B]{}\HSKeyword{record}\;\HSCon{ASTTypes}\;\mathbin{\HSCon{:}}\;\HSCon{Set}\;\HSKeyword{where}{}\<[E]%
\\
\>[B]{}\hsindent{3}{}\<[3]%
\>[3]{}\HSKeyword{field}{}\<[E]%
\\
\>[3]{}\hsindent{2}{}\<[5]%
\>[5]{}\HSCon{Input}\;{}\<[12]%
\>[12]{}\mathbin{\HSCon{:}}\;\HSCon{Set}{}\<[E]%
\\
\>[3]{}\hsindent{2}{}\<[5]%
\>[5]{}\HSCon{Output}\;\mathbin{\HSCon{:}}\;\HSSpecial{(}\HSCon{A}\;\mathbin{\HSCon{:}}\;\HSCon{Set}\HSSpecial{)}\;\mathrel{\HSSym{\to}} \;\HSCon{Set}{}\<[E]%
\\[\blanklineskip]%
\>[B]{}\hsindent{3}{}\<[3]%
\>[3]{}\HSCon{Exec}\;\mathbin{\HSCon{:}}\;\HSCon{Set}\;\mathrel{\HSSym{\to}} \;\HSCon{Set}{}\<[E]%
\\
\>[B]{}\hsindent{3}{}\<[3]%
\>[3]{}\HSCon{Exec}\;\HSCon{A}\;\mathrel{\HSSym{=}}\;\HSCon{Input}\;\mathrel{\HSSym{\to}} \;\HSCon{Output}\;\HSCon{A}{}\<[E]%
\ColumnHook
\end{hscode}\resethooks
  \end{subfigure}%
  \begin{subfigure}{0.50\linewidth}
    \begin{hscode}\SaveRestoreHook
\column{B}{@{}>{\hspre}l<{\hspost}@{}}%
\column{3}{@{}>{\hspre}l<{\hspost}@{}}%
\column{5}{@{}>{\hspre}l<{\hspost}@{}}%
\column{9}{@{}>{\hspre}l<{\hspost}@{}}%
\column{11}{@{}>{\hspre}l<{\hspost}@{}}%
\column{19}{@{}>{\hspre}l<{\hspost}@{}}%
\column{26}{@{}>{\hspre}l<{\hspost}@{}}%
\column{E}{@{}>{\hspre}l<{\hspost}@{}}%
\>[B]{}\HSKeyword{record}\;{}\<[9]%
\>[9]{}\HSCon{ASTOpSem}\;{}\<[19]%
\>[19]{}\HSSpecial{(}\HSCon{OP}\;\mathbin{\HSCon{:}}\;\HSCon{ASTOps}\HSSpecial{)}\;{}\<[E]%
\\
\>[9]{}\hsindent{2}{}\<[11]%
\>[11]{}\HSSpecial{(}\HSCon{Ty}\;\mathbin{\HSCon{:}}\;\HSCon{ASTTypes}\HSSpecial{)}\;\mathbin{\HSCon{:}}\;\HSCon{Set}\;\HSKeyword{where}{}\<[E]%
\\
\>[B]{}\hsindent{3}{}\<[3]%
\>[3]{}\HSKeyword{open}\;\HSCon{ASTTypes}\;\HSCon{Ty}{}\<[E]%
\\
\>[B]{}\hsindent{3}{}\<[3]%
\>[3]{}\HSKeyword{field}{}\<[E]%
\\
\>[3]{}\hsindent{2}{}\<[5]%
\>[5]{}\HSVar{runAST}\;\mathbin{\HSCon{:}}\;\HSSym{\forall}\;\HSSpecial{\HSSym{\{\mskip1.5mu} }\HSCon{A}\HSSpecial{\HSSym{\mskip1.5mu\}}}\;{}\<[26]%
\>[26]{}\mathrel{\HSSym{\to}} \;\HSCon{AST}\;\HSCon{OP}\;\HSCon{A}\;{}\<[E]%
\\
\>[26]{}\mathrel{\HSSym{\to}} \;\HSCon{Exec}\;\HSCon{A}{}\<[E]%
\ColumnHook
\end{hscode}\resethooks
  \end{subfigure}
  \caption{Operational semantics for an \ensuremath{\HSCon{AST}}}
  \label{fig:ast-opsem}
\end{figure}

\paragraph*{Operational Semantics}
Figure~\ref{fig:ast-opsem} shows what is required for \emph{running} \ensuremath{\HSCon{AST}}
programs. 
First, the user must supply the \ensuremath{\HSCon{Input}} type and
\ensuremath{\HSCon{Output}} type family (type argument \ensuremath{\HSCon{A}} is the type of values returned by the
program).
The type of \emph{executions} of a program producing values of type \ensuremath{\HSCon{A}} is a
function \ensuremath{\HSCon{Exec}\;\HSCon{A}} that maps an \ensuremath{\HSCon{Input}} to an \ensuremath{\HSCon{Output}\;\HSCon{A}}.
The user then provides an instance of \ensuremath{\HSCon{ASTOpSem}} for
the desired operations and types.
The only field of \ensuremath{\HSCon{ASTOpSem}} is \ensuremath{\HSVar{runAST}}, a function that transforms the AST of
an effectful computation producing values of type \ensuremath{\HSCon{A}} to a function of type
\ensuremath{\HSCon{Exec}\;\HSCon{A}}.

\begin{example}
  \label{ex:ast-rws-op}
  \begin{figure}[t]
    \centering
    \begin{hscode}\SaveRestoreHook
\column{B}{@{}>{\hspre}l<{\hspost}@{}}%
\column{3}{@{}>{\hspre}l<{\hspost}@{}}%
\column{14}{@{}>{\hspre}l<{\hspost}@{}}%
\column{16}{@{}>{\hspre}l<{\hspost}@{}}%
\column{24}{@{}>{\hspre}l<{\hspost}@{}}%
\column{28}{@{}>{\hspre}l<{\hspost}@{}}%
\column{E}{@{}>{\hspre}l<{\hspost}@{}}%
\>[B]{}\HSKeyword{data}\;\HSCon{RWSCmd}\;\HSSpecial{(}\HSCon{A}\;\mathbin{\HSCon{:}}\;\HSCon{Set}\HSSpecial{)}\;\mathbin{\HSCon{:}}\;\HSCon{Set}\;\HSKeyword{where}{}\<[E]%
\\
\>[B]{}\hsindent{3}{}\<[3]%
\>[3]{}\HSCon{RWSgets}\;\mathbin{\HSCon{:}}\;{}\<[14]%
\>[14]{}\HSSpecial{(}\HSVar{g}\;\mathbin{\HSCon{:}}\;\HSCon{St}\;\mathrel{\HSSym{\to}} \;\HSCon{A}\HSSpecial{)}\;\mathrel{\HSSym{\to}} \;\HSCon{RWSCmd}\;\HSCon{A}{}\<[E]%
\\
\>[B]{}\hsindent{3}{}\<[3]%
\>[3]{}\HSCon{RWSpass}\;\mathbin{\HSCon{:}}\;{}\<[14]%
\>[14]{}\HSCon{RWSCmd}\;\HSCon{A}{}\<[E]%
\\
\>[B]{}\hsindent{3}{}\<[3]%
\>[3]{}\HSVar{...}{}\<[E]%
\\
\>[B]{}\HSCon{RWSSubArg}\;\mathbin{\HSCon{:}}\;\HSSpecial{\HSSym{\{\mskip1.5mu} }\HSCon{A}\;\mathbin{\HSCon{:}}\;\HSCon{Set}\HSSpecial{\HSSym{\mskip1.5mu\}}}\;\HSSpecial{(}\HSVar{c}\;\mathbin{\HSCon{:}}\;\HSCon{RWSCmd}\;\HSCon{A}\HSSpecial{)}\;\mathrel{\HSSym{\to}} \;\HSCon{Set}{}\<[E]%
\\
\>[B]{}\HSCon{RWSSubArg}\;\HSSpecial{(}\HSCon{RWSgets}\;\HSVar{g}\HSSpecial{)}\;{}\<[24]%
\>[24]{}\mathrel{\HSSym{=}}\;\HSCon{Void}{}\<[E]%
\\
\>[B]{}\HSCon{RWSSubArg}\;\HSCon{RWSpass}\;{}\<[24]%
\>[24]{}\mathrel{\HSSym{=}}\;\HSCon{Unit}{}\<[E]%
\\
\>[B]{}\HSVar{...}{}\<[E]%
\\
\>[B]{}\HSCon{RWSSubRet}\;\mathbin{\HSCon{:}}\;\HSSpecial{\HSSym{\{\mskip1.5mu} }\HSCon{A}\;\mathbin{\HSCon{:}}\;\HSCon{Set}\HSSpecial{\HSSym{\mskip1.5mu\}}}\;\HSSpecial{\HSSym{\{\mskip1.5mu} }\HSVar{c}\;\mathbin{\HSCon{:}}\;\HSCon{RWSCmd}\;\HSCon{A}\HSSpecial{\HSSym{\mskip1.5mu\}}}\;\HSSpecial{(}\HSVar{r}\;\mathbin{\HSCon{:}}\;\HSCon{RWSSubArg}\;\HSVar{c}\HSSpecial{)}\;\mathrel{\HSSym{\to}} \;\HSCon{Set}{}\<[E]%
\\
\>[B]{}\HSCon{RWSSubRet}\;\HSSpecial{\HSSym{\{\mskip1.5mu} }\mathbin{\HSSym{\anonymous}} \HSSpecial{\HSSym{\mskip1.5mu\}}}\;\HSSpecial{\HSSym{\{\mskip1.5mu} }\HSCon{RWSgets}\;\HSVar{g}\HSSpecial{\HSSym{\mskip1.5mu\}}}\;{}\<[28]%
\>[28]{}\HSSpecial{(}\HSSpecial{)}{}\<[E]%
\\
\>[B]{}\HSCon{RWSSubRet}\;\HSSpecial{\HSSym{\{\mskip1.5mu} }\HSCon{A}\HSSpecial{\HSSym{\mskip1.5mu\}}}\;\HSSpecial{\HSSym{\{\mskip1.5mu} }\HSCon{RWSpass}\HSSpecial{\HSSym{\mskip1.5mu\}}}\;{}\<[28]%
\>[28]{}\mathbin{\HSSym{\anonymous}} \;\mathrel{\HSSym{=}}\;\HSCon{A}\;\mathbin{\HT{\times}}\;\HSSpecial{(}\HSCon{List}\;\HSCon{Wr}\;\mathrel{\HSSym{\to}} \;\HSCon{List}\;\HSCon{Wr}\HSSpecial{)}{}\<[E]%
\\
\>[B]{}\HSVar{...}{}\<[E]%
\\
\>[B]{}\HSCon{RWSOps}\;\mathbin{\HSCon{:}}\;\HSCon{ASTOps}{}\<[E]%
\\
\>[B]{}\HSCon{Cmd}\;\HSCon{RWSOps}\;{}\<[16]%
\>[16]{}\mathrel{\HSSym{=}}\;\HSCon{RWSCmd}{}\<[E]%
\\
\>[B]{}\HSCon{SubArg}\;\HSCon{RWSOps}\;{}\<[16]%
\>[16]{}\mathrel{\HSSym{=}}\;\HSCon{RWSSubArg}{}\<[E]%
\\
\>[B]{}\HSCon{SubRet}\;\HSCon{RWSOps}\;{}\<[16]%
\>[16]{}\mathrel{\HSSym{=}}\;\HSCon{RWSSubRet}{}\<[E]%
\ColumnHook
\end{hscode}\resethooks
    \caption{Commands and operational types for \ensuremath{\HSCon{RWS}}}
    \label{fig:ast-rws-op}
  \end{figure}

  Figure~\ref{fig:ast-rws-op} sketches the instantiation of \ensuremath{\HSCon{ASTOps}} for
  modeling \ensuremath{\HSCon{RWS}} programs.
  Due to space constraints, we only show two operations and omit the (straightforward)
  definition of the operational semantics; for the full definition,
  see module \ensuremath{\HSCon{Dijkstra.AST.RWS}}.
  \ensuremath{\HSCon{RWS}} operations \ensuremath{\HSVar{gets}\;\mathbin{\HSCon{:}}\;\HSSym{\forall}\;\HSSpecial{\HSSym{\{\mskip1.5mu} }\HSCon{A}\HSSpecial{\HSSym{\mskip1.5mu\}}}\;\mathrel{\HSSym{\to}} \;\HSSpecial{(}\HSCon{St}\;\mathrel{\HSSym{\to}} \;\HSCon{A}\HSSpecial{)}\;\mathrel{\HSSym{\to}} \;\HSCon{A}} and \ensuremath{\HSVar{pass}\;\mathbin{\HSCon{:}}\;\HSSym{\forall}\;\HSSpecial{\HSSym{\{\mskip1.5mu} }\HSCon{A}\HSSpecial{\HSSym{\mskip1.5mu\}}}\;\mathrel{\HSSym{\to}} \;\HSCon{RWS}\;\HSSpecial{(}\HSCon{A}\;\mathbin{\HT{\times}}\;\HSSpecial{(}\HSCon{List}\;\HSCon{Wr}\;\mathrel{\HSSym{\to}} \;\HSCon{List}\;\HSCon{Wr}\HSSpecial{)}\HSSpecial{)}\;\mathrel{\HSSym{\to}} \;\HSCon{RWS}\;\HSCon{A}} are modeled with constructors
  \ensuremath{\HSCon{RWSgets}} and \ensuremath{\HSCon{RWSpass}}.
  \ensuremath{\HSCon{RWSgets}} has no \ensuremath{\HSCon{RWS}} subcomputations, so the arity (given by
  \ensuremath{\HSCon{RWSubArg}\;\HSSpecial{(}\HSCon{RWSgets}\;\HSVar{g}\HSSpecial{)}}) is \ensuremath{\HSCon{Void}} (the empty type).
  \ensuremath{\HSCon{RWSpass}} has a single subcomputation (so the arity is \ensuremath{\HSCon{Unit}}); the type it
  returns is \ensuremath{\HSCon{A}\;\mathbin{\HT{\times}}\;\HSSpecial{(}\HSCon{List}\;\HSCon{Wr}\;\mathrel{\HSSym{\to}} \;\HSCon{List}\;\HSCon{Wr}\HSSpecial{)}}, where \ensuremath{\HSCon{A}} is the type of the entire
  \ensuremath{\HSCon{RWSpass}} computation.
  The \ensuremath{\HSCon{ASTTypes}} instance for \ensuremath{\HSCon{RWS}} (not shown)
  sets the fields to the definitions of \ensuremath{\HSCon{Input}} and \ensuremath{\HSCon{Output}} given in
  Section~\ref{sec:proof-engineering-pts}.
\end{example}

\subsection{Predicate Transformer Semantics}
\label{sec:ast-pts}

\begin{figure}[t]
  \begin{hscode}\SaveRestoreHook
\column{B}{@{}>{\hspre}l<{\hspost}@{}}%
\column{3}{@{}>{\hspre}l<{\hspost}@{}}%
\column{5}{@{}>{\hspre}l<{\hspost}@{}}%
\column{14}{@{}>{\hspre}l<{\hspost}@{}}%
\column{15}{@{}>{\hspre}l<{\hspost}@{}}%
\column{16}{@{}>{\hspre}l<{\hspost}@{}}%
\column{17}{@{}>{\hspre}l<{\hspost}@{}}%
\column{32}{@{}>{\hspre}l<{\hspost}@{}}%
\column{40}{@{}>{\hspre}l<{\hspost}@{}}%
\column{E}{@{}>{\hspre}l<{\hspost}@{}}%
\>[B]{}\HSKeyword{record}\;\HSCon{ASTTypes}\;\mathbin{\HSCon{:}}\;\HSCon{Set}\;\HSKeyword{where}{}\<[E]%
\\
\>[B]{}\hsindent{3}{}\<[3]%
\>[3]{}\HSVar{...}{}\<[E]%
\\
\>[B]{}\hsindent{3}{}\<[3]%
\>[3]{}\HSCon{Pre}\;{}\<[16]%
\>[16]{}\mathrel{\HSSym{=}}\;\HSCon{Input}\;\mathrel{\HSSym{\to}} \;\HSCon{Set}{}\<[E]%
\\
\>[B]{}\hsindent{3}{}\<[3]%
\>[3]{}\HSCon{Post}\;\HSCon{A}\;{}\<[16]%
\>[16]{}\mathrel{\HSSym{=}}\;\HSCon{Output}\;\HSCon{A}\;\mathrel{\HSSym{\to}} \;\HSCon{Set}{}\<[E]%
\\
\>[B]{}\hsindent{3}{}\<[3]%
\>[3]{}\HSCon{PredTrans}\;\HSCon{A}\;{}\<[16]%
\>[16]{}\mathrel{\HSSym{=}}\;\HSCon{Post}\;\HSCon{A}\;\mathrel{\HSSym{\to}} \;\HSCon{Pre}{}\<[E]%
\\[\blanklineskip]%
\>[B]{}\HSKeyword{record}\;\HSCon{ASTPredTrans}\;\HSSpecial{(}\HSCon{OP}\;\mathbin{\HSCon{:}}\;\HSCon{ASTOps}\HSSpecial{)}\;\HSSpecial{(}\HSCon{Ty}\;\mathbin{\HSCon{:}}\;\HSCon{ASTTypes}\HSSpecial{)}\;\mathbin{\HSCon{:}}\;\HSCon{Set}\;\HSKeyword{where}{}\<[E]%
\\
\>[B]{}\hsindent{3}{}\<[3]%
\>[3]{}\HSKeyword{open}\;\HSVar{...}{}\<[14]%
\>[14]{}\HSComment{ -\! - opens to avoid explicit references are omitted hereafter}{}\<[E]%
\\
\>[B]{}\hsindent{3}{}\<[3]%
\>[3]{}\HSKeyword{field}{}\<[E]%
\\
\>[3]{}\hsindent{2}{}\<[5]%
\>[5]{}\HSVar{returnPT}\;{}\<[15]%
\>[15]{}\mathbin{\HSCon{:}}\;\HSSym{\forall}\;\HSSpecial{\HSSym{\{\mskip1.5mu} }\HSCon{A}\HSSpecial{\HSSym{\mskip1.5mu\}}}\;\mathrel{\HSSym{\to}} \;\HSCon{A}\;{}\<[40]%
\>[40]{}\mathrel{\HSSym{\to}} \;\HSCon{PredTrans}\;\HSCon{A}{}\<[E]%
\\
\>[3]{}\hsindent{2}{}\<[5]%
\>[5]{}\HSVar{bindPT}\;{}\<[15]%
\>[15]{}\mathbin{\HSCon{:}}\;\HSSym{\forall}\;\HSSpecial{\HSSym{\{\mskip1.5mu} }\HSCon{A}\;\HSCon{B}\HSSpecial{\HSSym{\mskip1.5mu\}}}\;\mathrel{\HSSym{\to}} \;\HSSpecial{(}\HSCon{A}\;\mathrel{\HSSym{\to}} \;\HSCon{PredTrans}\;\HSCon{B}\HSSpecial{)}\;\mathrel{\HSSym{\to}} \;\HSCon{Input}{}\<[E]%
\\
\>[15]{}\hsindent{2}{}\<[17]%
\>[17]{}\mathrel{\HSSym{\to}} \;\HSCon{Post}\;\HSCon{B}\;{}\<[40]%
\>[40]{}\mathrel{\HSSym{\to}} \;\HSCon{Post}\;\HSCon{A}{}\<[E]%
\\
\>[3]{}\hsindent{2}{}\<[5]%
\>[5]{}\HSVar{opPT}\;{}\<[15]%
\>[15]{}\mathbin{\HSCon{:}}\;\HSSym{\forall}\;\HSSpecial{\HSSym{\{\mskip1.5mu} }\HSCon{A}\HSSpecial{\HSSym{\mskip1.5mu\}}}\;\mathrel{\HSSym{\to}} \;\HSSpecial{(}\HSVar{c}\;\mathbin{\HSCon{:}}\;\HSCon{Cmd}\;\HSCon{OP}\;\HSCon{A}\HSSpecial{)}{}\<[E]%
\\
\>[15]{}\hsindent{2}{}\<[17]%
\>[17]{}\mathrel{\HSSym{\to}} \;\HSSpecial{(}\HSSpecial{(}\HSVar{r}\;\mathbin{\HSCon{:}}\;\HSCon{SubArg}\;\HSCon{OP}\;\HSVar{c}\HSSpecial{)}\;{}\<[40]%
\>[40]{}\mathrel{\HSSym{\to}} \;\HSCon{PredTrans}\;\HSSpecial{(}\HSCon{SubRet}\;\HSCon{OP}\;\HSVar{r}\HSSpecial{)}\HSSpecial{)}\;{}\<[E]%
\\
\>[40]{}\mathrel{\HSSym{\to}} \;\HSCon{PredTrans}\;\HSCon{A}{}\<[E]%
\\[\blanklineskip]%
\>[B]{}\hsindent{3}{}\<[3]%
\>[3]{}\HSVar{predTrans}\;\mathbin{\HSCon{:}}\;\HSSym{\forall}\;\HSSpecial{\HSSym{\{\mskip1.5mu} }\HSCon{A}\HSSpecial{\HSSym{\mskip1.5mu\}}}\;\mathrel{\HSSym{\to}} \;\HSCon{AST}\;\HSCon{OP}\;\HSCon{A}\;{}\<[40]%
\>[40]{}\mathrel{\HSSym{\to}} \;\HSCon{PredTrans}\;\HSCon{A}{}\<[E]%
\\
\>[B]{}\hsindent{3}{}\<[3]%
\>[3]{}\HSVar{predTrans}\;\HSSpecial{(}\HSCon{ASTreturn}\;\HSVar{x}\HSSpecial{)}\;\HSVar{\mathit{P}}\;\HSVar{i}\;{}\<[32]%
\>[32]{}\mathrel{\HSSym{=}}\;\HSVar{returnPT}\;\HSVar{x}\;\HSVar{\mathit{P}}\;\HSVar{i}{}\<[E]%
\\
\>[B]{}\hsindent{3}{}\<[3]%
\>[3]{}\HSVar{predTrans}\;\HSSpecial{(}\HSCon{ASTbind}\;\HSVar{m}\;\HSVar{f}\HSSpecial{)}\;\HSVar{\mathit{P}}\;\HSVar{i}\;{}\<[32]%
\>[32]{}\mathrel{\HSSym{=}}\;\HSVar{predTrans}\;\HSVar{m}\;\HSSpecial{(}\HSVar{bindPT}\;\HSSpecial{(}\HSVar{predTrans}\;\mathbin{\HSSym{\circ}}\;\HSVar{f}\HSSpecial{)}\;\HSVar{i}\;\HSVar{\mathit{P}}\HSSpecial{)}\;\HSVar{i}{}\<[E]%
\\
\>[B]{}\hsindent{3}{}\<[3]%
\>[3]{}\HSVar{predTrans}\;\HSSpecial{(}\HSCon{ASTop}\;\HSVar{c}\;\HSVar{f}\HSSpecial{)}\;\HSVar{\mathit{P}}\;\HSVar{i}\;{}\<[32]%
\>[32]{}\mathrel{\HSSym{=}}\;\HSVar{opPT}\;\HSVar{c}\;\HSSpecial{(}\HSVar{predTrans}\;\mathbin{\HSSym{\circ}}\;\HSVar{f}\HSSpecial{)}\;\HSVar{\mathit{P}}\;\HSVar{i}{}\<[E]%
\ColumnHook
\end{hscode}\resethooks
  \caption{Predicate transformer semantics for an \ensuremath{\HSCon{AST}}}
  \label{fig:ast-pts}
\end{figure}

We can now define what it means to give a PTS to an \ensuremath{\HSCon{AST}}.
It is at \emph{this step} where the proof engineer instantiating the framework
decides what proof obligations are generated for effectful operations and
sequencing; we will see this more concretely in Example~\ref{ex:ast-rws-pts}.
The step that follows, described in Section~\ref{sec:ast-suff}, requires showing
that PTS \emph{agree} with the operational semantics. 

In Figure~\ref{fig:ast-pts}, we continue the listing of the record
\ensuremath{\HSCon{ASTTypes}}, which includes some definitions for convenience:
\ensuremath{\HSCon{Pre}} and \ensuremath{\HSCon{Post}} are the definitions of preconditions and
postconditions, and \ensuremath{\HSCon{PredTrans}} defines a predicate transformer as a function
that produces preconditions from postconditions.
A user wishing to assign a PTS to a particular choice of AST operations \ensuremath{\HSCon{OP}} and
input and output types \ensuremath{\HSCon{Ty}} provides three items, expressed as the three fields
of \ensuremath{\HSCon{ASTPredTrans}}.

Each field  of \ensuremath{\HSCon{ASTPredTrans}} corresponds to a clause of the definition of
\ensuremath{\HSVar{predTrans}}, the function that assigns a predicate transformer to \ensuremath{\HSCon{AST}\;\HSCon{OP}} programs.
Field \ensuremath{\HSVar{returnPT}} is a family of predicate transformers for \ensuremath{\HSCon{ASTreturn}}.
Field \ensuremath{\HSVar{bindPT}} is for composite operations of the form \ensuremath{\HSCon{ASTbind}\;\HSVar{m}\;\HSVar{f}\;\mathbin{\HSCon{:}}\;\HSCon{AST}\;\HSCon{OP}\;\HSCon{B}}.
Its purpose is to take a postcondition \ensuremath{\HSVar{\mathit{P}}\;\mathbin{\HSCon{:}}\;\HSCon{Post}\;\HSCon{B}} for the entire computation
and rephrase it as a postcondition (of type \ensuremath{\HSCon{Post}\;\HSCon{A}}) for \ensuremath{\HSVar{m}\;\mathbin{\HSCon{:}}\;\HSCon{AST}\;\HSCon{OP}\;\HSCon{A}}.
The definition of \ensuremath{\HSVar{bindPT}} supplied by the user should use the assumed
family of predicate transformers (the argument of type \ensuremath{\HSCon{A}\;\mathrel{\HSSym{\to}} \;\HSCon{PredTrans}\;\HSCon{B}}) to
express a sufficient precondition of \ensuremath{\HSVar{f}} to prove \ensuremath{\HSVar{\mathit{P}}}, and then make that
precondition the \emph{postcondition} for \ensuremath{\HSVar{m}} (see
Example~\ref{ex:ast-rws-pts}). 
Field \ensuremath{\HSVar{opPT}} gives a predicate transformer for every command \ensuremath{\HSVar{c}\;\mathbin{\HSCon{:}}\;\HSCon{Cmd}\;\HSCon{OP}\;\HSCon{A}}, given a family of predicate transformers for the subcomputations (if any)
given as arguments to that command.

Given these pieces, the function \ensuremath{\HSVar{predTrans}} that assigns a predicate
transformer to every \ensuremath{\HSVar{m}\;\mathbin{\HSCon{:}}\;\HSCon{AST}\;\HSCon{OP}\;\HSCon{A}} is defined by induction over \ensuremath{\HSVar{m}}.
We again call attention to the fact that, because our \ensuremath{\HSCon{AST}} type has the
constructor \ensuremath{\HSCon{ASTbind}}, users can tailor a convenient predicate transformer to
assign to composite operations directly, rather than requiring an explicit
compositionality lemma.

\begin{example}
  \label{ex:ast-rws-pts}
  \begin{figure}[t]
    \begin{hscode}\SaveRestoreHook
\column{B}{@{}>{\hspre}l<{\hspost}@{}}%
\column{3}{@{}>{\hspre}l<{\hspost}@{}}%
\column{11}{@{}>{\hspre}l<{\hspost}@{}}%
\column{20}{@{}>{\hspre}l<{\hspost}@{}}%
\column{33}{@{}>{\hspre}l<{\hspost}@{}}%
\column{E}{@{}>{\hspre}l<{\hspost}@{}}%
\>[B]{}\HSCon{RWSbindPost}\;\mathbin{\HSCon{:}}\;\HSSpecial{(}\HSVar{outs}\;\mathbin{\HSCon{:}}\;\HSCon{List}\;\HSCon{Wr}\HSSpecial{)}\;\HSSpecial{\HSSym{\{\mskip1.5mu} }\HSCon{A}\;\mathbin{\HSCon{:}}\;\HSCon{Set}\HSSpecial{\HSSym{\mskip1.5mu\}}}\;\mathrel{\HSSym{\to}} \;\HSCon{Post}\;\HSCon{A}\;\mathrel{\HSSym{\to}} \;\HSCon{Post}\;\HSCon{A}{}\<[E]%
\\
\>[B]{}\HSCon{RWSbindPost}\;\HSVar{outs}\;\HSVar{\mathit{P}}\;\HSSpecial{(}\HSVar{x}\;\mathbin{\HSSym{,}}\;\HSVar{st}\;\mathbin{\HSSym{,}}\;\HSVar{outs'}\HSSpecial{)}\;\mathrel{\HSSym{=}}\;\HSVar{\mathit{P}}\;\HSSpecial{(}\HSVar{x}\;\mathbin{\HSSym{,}}\;\HSVar{st}\;\mathbin{\HSSym{,}}\;\HSVar{outs}\;\mathbin{\HSSym{\plus}} \;\HSVar{outs'}\HSSpecial{)}{}\<[E]%
\\[\blanklineskip]%
\>[B]{}\HSCon{RWSpassPost}\;\mathbin{\HSCon{:}}\;\HSSym{\forall}\;\HSSpecial{\HSSym{\{\mskip1.5mu} }\HSCon{A}\HSSpecial{\HSSym{\mskip1.5mu\}}}\;\mathrel{\HSSym{\to}} \;\HSCon{Post}\;\HSCon{A}\;\mathrel{\HSSym{\to}} \;\HSCon{Post}\;\HSSpecial{(}\HSCon{A}\;\mathbin{\HT{\times}}\;\HSSpecial{(}\HSCon{List}\;\HSCon{Wr}\;\mathrel{\HSSym{\to}} \;\HSCon{List}\;\HSCon{Wr}\HSSpecial{)}\HSSpecial{)}{}\<[E]%
\\
\>[B]{}\HSCon{RWSpassPost}\;\HSVar{\mathit{P}}\;\HSSpecial{(}\HSSpecial{(}\HSVar{x}\;\mathbin{\HSSym{,}}\;\HSVar{f}\HSSpecial{)}\;\mathbin{\HSSym{,}}\;\HSVar{s}\;\mathbin{\HSSym{,}}\;\HSVar{o}\HSSpecial{)}\;\mathrel{\HSSym{=}}\;\HSSym{\forall}\;\HSVar{o'}\;\mathrel{\HSSym{\to}} \;\HSVar{o'}\;\mathbin{\HT{\equiv}}\;\HSVar{f}\;\HSVar{o}\;\mathrel{\HSSym{\to}} \;\HSVar{\mathit{P}}\;\HSSpecial{(}\HSVar{x}\;\mathbin{\HSSym{,}}\;\HSVar{s}\;\mathbin{\HSSym{,}}\;\HSVar{o'}\HSSpecial{)}{}\<[E]%
\\[\blanklineskip]%
\>[B]{}\HSCon{RWSPT}\;\mathbin{\HSCon{:}}\;\HSCon{ASTPredTrans}\;\HSCon{RWSOps}\;\HSCon{RWSTypes}{}\<[E]%
\\
\>[B]{}\HSVar{returnPT}\;{}\<[11]%
\>[11]{}\HSCon{RWSPT}\;\HSVar{x}\;\HSVar{\mathit{P}}\;\HSSpecial{(}\HSVar{e}\;\mathbin{\HSSym{,}}\;\HSVar{s}\HSSpecial{)}\;\mathrel{\HSSym{=}}\;\HSVar{\mathit{P}}\;\HSSpecial{(}\HSVar{x}\;\mathbin{\HSSym{,}}\;\HSVar{st,}\;\HSVar{[]}\HSSpecial{)}{}\<[E]%
\\
\>[B]{}\HSVar{bindPT}\;{}\<[11]%
\>[11]{}\HSCon{RWSPT}\;\HSVar{f}\;\HSSpecial{(}\HSVar{e}\;\mathbin{\HSSym{,}}\;\HSVar{\mathit{s}_{0}}\HSSpecial{)}\;\HSVar{\mathit{P}}\;\HSSpecial{(}\HSVar{x}\;\mathbin{\HSSym{,}}\;\HSVar{\mathit{s}_{1}}\;\mathbin{\HSSym{,}}\;\HSVar{o}\HSSpecial{)}\;\mathrel{\HSSym{=}}{}\<[E]%
\\
\>[B]{}\hsindent{3}{}\<[3]%
\>[3]{}\HSSym{\forall}\;\HSVar{r}\;\mathrel{\HSSym{\to}} \;\HSVar{r}\;\mathbin{\HT{\equiv}}\;\HSVar{x}\;\mathrel{\HSSym{\to}} \;\HSVar{f}\;\HSVar{r}\;\HSSpecial{(}\HSCon{RWSbindPost}\;\HSVar{o}\;\HSVar{\mathit{P}}\HSSpecial{)}\;\HSSpecial{(}\HSVar{e}\;\mathbin{\HSSym{,}}\;\HSVar{\mathit{s}_{1}}\HSSpecial{)}{}\<[E]%
\\
\>[B]{}\HSVar{opPT}\;{}\<[11]%
\>[11]{}\HSCon{RWSPT}\;{}\<[20]%
\>[20]{}\HSSpecial{(}\HSCon{RWSgets}\;\HSVar{g}\HSSpecial{)}\;{}\<[33]%
\>[33]{}\HSVar{f}\;\HSVar{\mathit{P}}\;\HSSpecial{(}\HSVar{e}\;\mathbin{\HSSym{,}}\;\HSVar{s}\HSSpecial{)}\;\mathrel{\HSSym{=}}\;\HSVar{\mathit{P}}\;\HSSpecial{(}\HSVar{g}\;\HSVar{s}\;\mathbin{\HSSym{,}}\;\HSVar{s}\;\mathbin{\HSSym{,}}\;\HSVar{[]}\HSSpecial{)}{}\<[E]%
\\
\>[B]{}\HSVar{opPT}\;{}\<[11]%
\>[11]{}\HSCon{RWSPT}\;\HSSpecial{\HSSym{\{\mskip1.5mu} }\HSCon{A}\HSSpecial{\HSSym{\mskip1.5mu\}}}\;\HSCon{RWSpass}\;{}\<[33]%
\>[33]{}\HSVar{f}\;\HSVar{\mathit{P}}\;\HSSpecial{(}\HSVar{e}\;\mathbin{\HSSym{,}}\;\HSVar{s}\HSSpecial{)}\;\mathrel{\HSSym{=}}\;\HSVar{f}\;\HT{unit}\;\HSSpecial{(}\HSCon{RWSpassPost}\;\HSVar{\mathit{P}}\HSSpecial{)}\;\HSSpecial{(}\HSVar{e}\;\mathbin{\HSSym{,}}\;\HSVar{s}\HSSpecial{)}{}\<[E]%
\ColumnHook
\end{hscode}\resethooks
    \caption{Predicate transformer semantics for \ensuremath{\HSCon{RWS}}}
    \label{fig:ast-rws-pts}
  \end{figure}

  Figure~\ref{fig:ast-rws-pts} shows the instantiation of \ensuremath{\HSCon{ASTPredTrans}}
  (omitting operations other than \ensuremath{\HSCon{RWSgets}} and \ensuremath{\HSCon{RWSpass}}).
  For \ensuremath{\HSVar{returnPT}}, the precondition we return is that the postcondition \ensuremath{\HSVar{\mathit{P}}}
  holds for the returned value, the current state, and an empty list of messages
  (there are no state changes or messages emitted).
  The \emph{post}condition we return for \ensuremath{\HSVar{bindPT}} is trickier: \ensuremath{\HSVar{\mathit{P}}} is the
  postcondition we wish to hold for \ensuremath{\HSCon{ASTbind}\;\HSVar{\mathit{m}_{1}}\;\HSVar{\mathit{m}_{2}}}, and what we return is the
  postcondition that should hold of \ensuremath{\HSVar{\mathit{m}_{1}}} to establish this.
  Because \ensuremath{\HSVar{x}} (the result of executing \ensuremath{\HSVar{\mathit{m}_{1}}}) may be instantiated to an unwieldy
  expression, we alias \ensuremath{\HSVar{x}} as \ensuremath{\HSVar{r}} and give this to the predicate transformer \ensuremath{\HSVar{f}}
  that is assigned to \ensuremath{\HSVar{\mathit{m}_{2}}}.
  We also have that \ensuremath{\HSVar{\mathit{m}_{1}}} emitted \ensuremath{\HSVar{o}} as output, so we use \ensuremath{\HSCon{RWSbindPost}} to express
  that the postcondition should hold for the result of appending these to the
  emitted messages of \ensuremath{\HSVar{\mathit{m}_{2}}}.

  We treat \ensuremath{\HSCon{RWSgets}} similarly to \ensuremath{\HSVar{returnPT}}: we require the postcondition to
  hold of \ensuremath{\HSVar{g}\;\HSVar{s}}, where \ensuremath{\HSVar{g}} is the user-supplied getter function.
  For \ensuremath{\HSCon{RWSpass}}, we apply the predicate transformer \ensuremath{\HSVar{f}} assigned to the
  subcomputation (think \ensuremath{\HSVar{inner}} from Section~\ref{sec:proof-engineering-pts}) to
  \ensuremath{\HSCon{RWSpassPost}\;\HSVar{\mathit{P}}}, which says that postcondition \ensuremath{\HSVar{\mathit{P}}} holds when we modify the
  output of the subcomputation with the returned function \ensuremath{\HSVar{h}\;\mathbin{\HSCon{:}}\;\HSCon{List}\;\HSCon{Wr}\;\mathrel{\HSSym{\to}} \;\HSCon{List}\;\HSCon{Wr}}.

\end{example}

\subsection{Agreement of Semantics}
\label{sec:ast-suff}

In Sections~\ref{sec:ast} and \ref{sec:ast-pts}, we described
how to assign operational and predicate transformer semantics to a set of
effectful operations.
We now describe how to show that two such semantics \emph{agree}.
The obligations to show one direction of this agreement are
formalized by \ensuremath{\HSCon{ASTSufficientPT}}, shown in Figure~\ref{fig:ast-suff}.

\begin{figure}
  \centering
  \begin{hscode}\SaveRestoreHook
\column{B}{@{}>{\hspre}l<{\hspost}@{}}%
\column{3}{@{}>{\hspre}l<{\hspost}@{}}%
\column{5}{@{}>{\hspre}l<{\hspost}@{}}%
\column{16}{@{}>{\hspre}l<{\hspost}@{}}%
\column{19}{@{}>{\hspre}l<{\hspost}@{}}%
\column{31}{@{}>{\hspre}l<{\hspost}@{}}%
\column{E}{@{}>{\hspre}l<{\hspost}@{}}%
\>[B]{}\HSKeyword{record}\;\HSCon{ASTSufficientPT}\;\HSSpecial{\HSSym{\{\mskip1.5mu} }\HSCon{OP}\;\mathbin{\HSCon{:}}\;\HSCon{ASTOps}\HSSpecial{\HSSym{\mskip1.5mu\}}}\;\HSSpecial{\HSSym{\{\mskip1.5mu} }\HSCon{Ty}\;\mathbin{\HSCon{:}}\;\HSCon{ASTTypes}\HSSpecial{\HSSym{\mskip1.5mu\}}}{}\<[E]%
\\
\>[B]{}\hsindent{3}{}\<[3]%
\>[3]{}\HSSpecial{(}\HSCon{OpSem}\;\mathbin{\HSCon{:}}\;\HSCon{ASTOpSem}\;\HSCon{OP}\;\HSCon{Ty}\HSSpecial{)}\;\HSSpecial{(}\HSCon{PT}\;\mathbin{\HSCon{:}}\;\HSCon{ASTPredTrans}\;\HSCon{OP}\;\HSCon{Ty}\HSSpecial{)}\;\mathbin{\HSCon{:}}\;\HSCon{Set}\;\HSKeyword{where}{}\<[E]%
\\[\blanklineskip]%
\>[B]{}\hsindent{3}{}\<[3]%
\>[3]{}\HSCon{Sufficient}\;\mathbin{\HSCon{:}}\;\HSSpecial{(}\HSCon{A}\;\mathbin{\HSCon{:}}\;\HSCon{Set}\HSSpecial{)}\;\HSSpecial{(}\HSVar{m}\;\mathbin{\HSCon{:}}\;\HSCon{AST}\;\HSCon{OP}\;\HSCon{A}\HSSpecial{)}\;\mathrel{\HSSym{\to}} \;\HSCon{Set}{}\<[E]%
\\
\>[B]{}\hsindent{3}{}\<[3]%
\>[3]{}\HSCon{Sufficient}\;\HSCon{A}\;\HSVar{m}\;\mathrel{\HSSym{=}}\;\HSSym{\forall}\;\HSVar{\mathit{P}}\;\HSVar{i}\;\mathrel{\HSSym{\to}} \;\HSSpecial{(}\HSVar{wp}\;\mathbin{\HSCon{:}}\;\HSVar{predTrans}\;\HSVar{m}\;\HSVar{\mathit{P}}\;\HSVar{i}\HSSpecial{)}\;\mathrel{\HSSym{\to}} \;\HSVar{\mathit{P}}\;\HSSpecial{(}\HSVar{runAST}\;\HSVar{m}\;\HSVar{i}\HSSpecial{)}{}\<[E]%
\\[\blanklineskip]%
\>[B]{}\hsindent{3}{}\<[3]%
\>[3]{}\HSKeyword{field}{}\<[E]%
\\
\>[3]{}\hsindent{2}{}\<[5]%
\>[5]{}\HSVar{returnSuf}\;{}\<[16]%
\>[16]{}\mathbin{\HSCon{:}}\;{}\<[19]%
\>[19]{}\HSSym{\forall}\;\HSSpecial{\HSSym{\{\mskip1.5mu} }\HSCon{A}\HSSpecial{\HSSym{\mskip1.5mu\}}}\;\HSVar{x}\;\mathrel{\HSSym{\to}} \;\HSCon{Sufficient}\;\HSCon{A}\;\HSSpecial{(}\HSCon{ASTreturn}\;\HSVar{x}\HSSpecial{)}{}\<[E]%
\\
\>[3]{}\hsindent{2}{}\<[5]%
\>[5]{}\HSVar{bindSuf}\;{}\<[16]%
\>[16]{}\mathbin{\HSCon{:}}\;{}\<[19]%
\>[19]{}\HSSym{\forall}\;\HSSpecial{\HSSym{\{\mskip1.5mu} }\HSCon{A}\;\HSCon{B}\HSSpecial{\HSSym{\mskip1.5mu\}}}\;\HSSpecial{(}\HSVar{m}\;\mathbin{\HSCon{:}}\;\HSCon{AST}\;\HSCon{OP}\;\HSCon{A}\HSSpecial{)}\;\HSSpecial{(}\HSVar{f}\;\mathbin{\HSCon{:}}\;\HSCon{A}\;\mathrel{\HSSym{\to}} \;\HSCon{AST}\;\HSCon{OP}\;\HSCon{B}\HSSpecial{)}{}\<[E]%
\\
\>[19]{}\mathrel{\HSSym{\to}} \;\HSCon{Sufficient}\;\HSCon{A}\;\HSVar{m}\;\mathrel{\HSSym{\to}} \;\HSSpecial{(}\HSSym{\forall}\;\HSVar{x}\;\mathrel{\HSSym{\to}} \;\HSCon{Sufficient}\;\HSCon{B}\;\HSSpecial{(}\HSVar{f}\;\HSVar{x}\HSSpecial{)}\HSSpecial{)}{}\<[E]%
\\
\>[19]{}\mathrel{\HSSym{\to}} \;\HSCon{Sufficient}\;\HSCon{B}\;\HSSpecial{(}\HSCon{ASTbind}\;\HSVar{m}\;\HSVar{f}\HSSpecial{)}{}\<[E]%
\\
\>[3]{}\hsindent{2}{}\<[5]%
\>[5]{}\HSVar{opSuf}\;{}\<[16]%
\>[16]{}\mathbin{\HSCon{:}}\;{}\<[19]%
\>[19]{}\HSSym{\forall}\;\HSSpecial{\HSSym{\{\mskip1.5mu} }\HSCon{A}\HSSpecial{\HSSym{\mskip1.5mu\}}}\;{}\<[31]%
\>[31]{}\mathrel{\HSSym{\to}} \;\HSSpecial{(}\HSVar{c}\;\mathbin{\HSCon{:}}\;\HSCon{Cmd}\;\HSCon{OP}\;\HSCon{A}\HSSpecial{)}\;\HSSpecial{(}\HSVar{f}\;\mathbin{\HSCon{:}}\;\HSCon{SubArg}\;\HSCon{OP}\;\HSVar{c}\;\mathrel{\HSSym{\to}} \;\HSCon{AST}\;\HSCon{OP}\;\HSSpecial{(}\HSCon{SubRet}\;\HSCon{OP}\;\HSVar{c}\HSSpecial{)}\HSSpecial{)}\;{}\<[E]%
\\
\>[31]{}\mathrel{\HSSym{\to}} \;\HSSpecial{(}\HSSym{\forall}\;\HSVar{r}\;\mathrel{\HSSym{\to}} \;\HSCon{Sufficient}\;\HSSpecial{(}\HSCon{SubRet}\;\HSCon{OP}\;\HSVar{c}\HSSpecial{)}\;\HSSpecial{(}\HSVar{f}\;\HSVar{r}\HSSpecial{)}\HSSpecial{)}\;{}\<[E]%
\\
\>[31]{}\mathrel{\HSSym{\to}} \;\HSCon{Sufficient}\;\HSCon{A}\;\HSSpecial{(}\HSCon{ASTop}\;\HSVar{c}\;\HSVar{f}\HSSpecial{)}{}\<[E]%
\\[\blanklineskip]%
\>[B]{}\hsindent{3}{}\<[3]%
\>[3]{}\HSVar{sufficient}\;\mathbin{\HSCon{:}}\;\HSSym{\forall}\;\HSSpecial{\HSSym{\{\mskip1.5mu} }\HSCon{A}\HSSpecial{\HSSym{\mskip1.5mu\}}}\;\mathrel{\HSSym{\to}} \;\HSSpecial{(}\HSVar{m}\;\mathbin{\HSCon{:}}\;\HSCon{AST}\;\HSCon{OP}\;\HSCon{A}\HSSpecial{)}\;\mathrel{\HSSym{\to}} \;\HSCon{Sufficient}\;\HSCon{A}\;\HSVar{m}{}\<[E]%
\\
\>[B]{}\hsindent{3}{}\<[3]%
\>[3]{}\HSVar{sufficient}\;\mathrel{\HSSym{=}}\;\HSVar{...}{}\<[E]%
\ColumnHook
\end{hscode}\resethooks
  \caption{Sufficiency lemmas for operational semantics and PTS}
  \label{fig:ast-suff}
\end{figure}

\ensuremath{\HSCon{Sufficient}} says that,
for a given effectful program \ensuremath{\HSVar{m}\;\mathbin{\HSCon{:}}\;\HSCon{AST}\;\HSCon{OP}\;\HSCon{A}}, the predicate transformer
\ensuremath{\HSVar{predTrans}\;\HSVar{m}} returns, for every postcondition \ensuremath{\HSVar{\mathit{P}}}, a precondition
\emph{sufficient} for proving that, for any input \ensuremath{\HSVar{i}}, \ensuremath{\HSVar{\mathit{P}}} is true of the result
obtained from running \ensuremath{\HSVar{m}} with input \ensuremath{\HSVar{i}} using the operational semantics;
henceforth we abbreviate this and say that \emph{the predicate transformer for
  \ensuremath{\HSVar{m}} is sufficient}.
To prove sufficiency, our framework imposes three obligations on the user,
corresponding to the three constructors of \ensuremath{\HSCon{AST}}; they are as follows.
\begin{enumerate}
\item Instantiating the field \ensuremath{\HSVar{returnSuf}} requires that the predicate
  transformer corresponding to \ensuremath{\HSCon{ASTreturn}\;\HSVar{x}} (for any \ensuremath{\HSVar{x}\;\mathbin{\HSCon{:}}\;\HSCon{A}}) is sufficient. 

\item For field \ensuremath{\HSVar{bindSuf}}, the user assumes that the predicate transformers
  assigned to \ensuremath{\HSVar{m}} and (all instances of) \ensuremath{\HSVar{f}} are sufficient, and must prove that
  the predicate transformer obtained from \ensuremath{\HSCon{ASTbind}\;\HSVar{m}\;\HSVar{f}} is sufficient.
  
\item Finally, for field \ensuremath{\HSVar{opSuf}}, assuming that, for an arbitrary command
  \ensuremath{\HSVar{c}} and subcomputation \ensuremath{\HSVar{f}}, the predicate transformer obtained from the result
  of running \ensuremath{\HSVar{f}} with any possible response value is sufficient, the user must
  show that the predicate transformer for \ensuremath{\HSCon{ASTop}\;\HSVar{c}\;\HSVar{f}} is sufficient.
\end{enumerate}

\noindent With these three obligations met, the proof of sufficiency
(\ensuremath{\HSVar{sufficient}}) proceeds by a straightforward induction.

The other direction of agreement is captured by \ensuremath{\HSCon{Necessary}} (not shown), which says that,
for an effectful program \ensuremath{\HSVar{m}\;\mathbin{\HSCon{:}}\;\HSCon{AST}\;\HSCon{OP}\;\HSCon{A}}, for every postcondition \ensuremath{\HSVar{\mathit{P}}} and
input \ensuremath{\HSVar{i}}, if the output achieved by running \ensuremath{\HSVar{m}} on \ensuremath{\HSVar{i}} satisfies \ensuremath{\HSVar{\mathit{P}}}, then \ensuremath{\HSVar{i}}
satisfies the precondition returned by \ensuremath{\HSVar{predTrans}\;\HSVar{m}\;\HSVar{\mathit{P}}}.
Similar to \ensuremath{\HSCon{Sufficient}}, the framework imposes an obligation on the user for
each constructor of \ensuremath{\HSCon{AST}}, and uses these to prove \ensuremath{\HSCon{Necessary}}; details can be
found in module \ensuremath{\HSCon{Dijstra.AST.Core}}.

\section{Generic Branching Operations}
\label{sec:branching}

Branching operations may be used \emph{in} effectful code, but do not
\emph{themselves} have effects.
Therefore, our framework can \emph{generically} extend any
set of commands and their predicate transformer semantics to include a few
common branching operations (see module \ensuremath{\HSCon{Dijkstra.AST.Core}}), re-expressing their proof obligations
to avoid the issues outlined in
Section~\ref{sec:prog-direct}.

\subsection{Branching Commands}
\label{sec:branch-cmd}

\begin{figure}[t]
  \begin{hscode}\SaveRestoreHook
\column{B}{@{}>{\hspre}l<{\hspost}@{}}%
\column{3}{@{}>{\hspre}l<{\hspost}@{}}%
\column{10}{@{}>{\hspre}l<{\hspost}@{}}%
\column{13}{@{}>{\hspre}l<{\hspost}@{}}%
\column{24}{@{}>{\hspre}l<{\hspost}@{}}%
\column{25}{@{}>{\hspre}l<{\hspost}@{}}%
\column{29}{@{}>{\hspre}l<{\hspost}@{}}%
\column{34}{@{}>{\hspre}l<{\hspost}@{}}%
\column{35}{@{}>{\hspre}l<{\hspost}@{}}%
\column{38}{@{}>{\hspre}l<{\hspost}@{}}%
\column{44}{@{}>{\hspre}l<{\hspost}@{}}%
\column{E}{@{}>{\hspre}l<{\hspost}@{}}%
\>[B]{}\HSKeyword{data}\;\HSCon{BranchCmd}\;\HSSpecial{(}\HSCon{A}\;\mathbin{\HSCon{:}}\;\HSCon{Set}\HSSpecial{)}\;\mathbin{\HSCon{:}}\;\HSCon{Set}\;\HSKeyword{where}{}\<[E]%
\\
\>[B]{}\hsindent{3}{}\<[3]%
\>[3]{}\HSCon{BCif}\;{}\<[13]%
\>[13]{}\mathbin{\HSCon{:}}\;\HSCon{Bool}\;{}\<[44]%
\>[44]{}\mathrel{\HSSym{\to}} \;\HSCon{BranchCmd}\;\HSCon{A}{}\<[E]%
\\
\>[B]{}\hsindent{3}{}\<[3]%
\>[3]{}\HSCon{BCeither}\;{}\<[13]%
\>[13]{}\mathbin{\HSCon{:}}\;\HSSym{\forall}\;\HSSpecial{\HSSym{\{\mskip1.5mu} }\HSCon{B}\;\HSCon{C}\HSSpecial{\HSSym{\mskip1.5mu\}}}\;{}\<[29]%
\>[29]{}\mathrel{\HSSym{\to}} \;\HSCon{Either}\;\HSCon{B}\;\HSCon{C}\;{}\<[44]%
\>[44]{}\mathrel{\HSSym{\to}} \;\HSCon{BranchCmd}\;\HSCon{A}{}\<[E]%
\\
\>[B]{}\hsindent{3}{}\<[3]%
\>[3]{}\HSCon{BCmaybe}\;{}\<[13]%
\>[13]{}\mathbin{\HSCon{:}}\;\HSSym{\forall}\;\HSSpecial{\HSSym{\{\mskip1.5mu} }\HSCon{B}\HSSpecial{\HSSym{\mskip1.5mu\}}}\;{}\<[29]%
\>[29]{}\mathrel{\HSSym{\to}} \;\HSCon{Maybe}\;\HSCon{B}\;{}\<[44]%
\>[44]{}\mathrel{\HSSym{\to}} \;\HSCon{BranchCmd}\;\HSCon{A}{}\<[E]%
\\[\blanklineskip]%
\>[B]{}\HSCon{BranchSubArg}\;\mathbin{\HSCon{:}}\;\HSSym{\forall}\;\HSSpecial{\HSSym{\{\mskip1.5mu} }\HSCon{A}\HSSpecial{\HSSym{\mskip1.5mu\}}}\;\mathrel{\HSSym{\to}} \;\HSCon{BranchCmd}\;\HSCon{A}\;\mathrel{\HSSym{\to}} \;\HSCon{Set}{}\<[E]%
\\
\>[B]{}\HSCon{BranchSubArg}\;\HSSpecial{(}\HSCon{BCif}\;{}\<[34]%
\>[34]{}\HSVar{x}\HSSpecial{)}\;{}\<[38]%
\>[38]{}\mathrel{\HSSym{=}}\;\HSCon{Bool}{}\<[E]%
\\
\>[B]{}\HSCon{BranchSubArg}\;\HSSpecial{(}\HSCon{BCeither}\;{}\<[25]%
\>[25]{}\HSSpecial{\HSSym{\{\mskip1.5mu} }\HSCon{B}\HSSpecial{\HSSym{\mskip1.5mu\}}}\;\HSSpecial{\HSSym{\{\mskip1.5mu} }\HSCon{C}\HSSpecial{\HSSym{\mskip1.5mu\}}}\;{}\<[34]%
\>[34]{}\HSVar{x}\HSSpecial{)}\;{}\<[38]%
\>[38]{}\mathrel{\HSSym{=}}\;\HSCon{Either}\;\HSCon{B}\;\HSCon{C}{}\<[E]%
\\
\>[B]{}\HSCon{BranchSubArg}\;\HSSpecial{(}\HSCon{BCmaybe}\;{}\<[25]%
\>[25]{}\HSSpecial{\HSSym{\{\mskip1.5mu} }\HSCon{B}\HSSpecial{\HSSym{\mskip1.5mu\}}}\;{}\<[34]%
\>[34]{}\HSVar{x}\HSSpecial{)}\;{}\<[38]%
\>[38]{}\mathrel{\HSSym{=}}\;\HSCon{Maybe}\;\HSCon{B}{}\<[E]%
\\[\blanklineskip]%
\>[B]{}\HSCon{BranchSubRet}\;\mathbin{\HSCon{:}}\;\HSSym{\forall}\;\HSSpecial{\HSSym{\{\mskip1.5mu} }\HSCon{A}\HSSpecial{\HSSym{\mskip1.5mu\}}}\;\HSSpecial{\HSSym{\{\mskip1.5mu} }\HSVar{c}\;\mathbin{\HSCon{:}}\;\HSCon{BranchCmd}\;\HSCon{A}\HSSpecial{\HSSym{\mskip1.5mu\}}}\;\mathrel{\HSSym{\to}} \;\HSCon{BranchSubArg}\;\HSVar{c}\;\mathrel{\HSSym{\to}} \;\HSCon{Set}{}\<[E]%
\\
\>[B]{}\HSCon{BranchSubRet}\;\HSSpecial{\HSSym{\{\mskip1.5mu} }\HSCon{A}\HSSpecial{\HSSym{\mskip1.5mu\}}}\;\mathbin{\HSSym{\anonymous}} \;\mathrel{\HSSym{=}}\;\HSCon{A}{}\<[E]%
\\[\blanklineskip]%
\>[B]{}\HSKeyword{module}\;\HSCon{ASTExtension}\;\HSSpecial{(}\HSCon{O}\;\mathbin{\HSCon{:}}\;\HSCon{ASTOps}\HSSpecial{)}\;\HSKeyword{where}{}\<[E]%
\\[\blanklineskip]%
\>[B]{}\hsindent{3}{}\<[3]%
\>[3]{}\HSCon{BranchOps}\;\mathbin{\HSCon{:}}\;\HSCon{ASTOps}{}\<[E]%
\\
\>[B]{}\hsindent{3}{}\<[3]%
\>[3]{}\HSCon{Cmd}\;{}\<[10]%
\>[10]{}\HSCon{BranchOps}\;\HSCon{A}\;{}\<[38]%
\>[38]{}\mathrel{\HSSym{=}}\;\HSCon{Either}\;\HSSpecial{(}\HSCon{Cmd}\;\HSCon{O}\;\HSCon{A}\HSSpecial{)}\;\HSSpecial{(}\HSCon{BranchCmd}\;\HSCon{A}\HSSpecial{)}{}\<[E]%
\\
\>[B]{}\hsindent{3}{}\<[3]%
\>[3]{}\HSCon{SubArg}\;\HSCon{BranchOps}\;{}\<[24]%
\>[24]{}\HSSpecial{(}\HT{left}\;\HSVar{x}\HSSpecial{)}\;{}\<[38]%
\>[38]{}\mathrel{\HSSym{=}}\;\HSCon{SubArg}\;\HSCon{O}\;\HSVar{x}{}\<[E]%
\\
\>[B]{}\hsindent{3}{}\<[3]%
\>[3]{}\HSCon{SubArg}\;\HSCon{BranchOps}\;{}\<[24]%
\>[24]{}\HSSpecial{(}\HT{right}\;\HSVar{y}\HSSpecial{)}\;{}\<[38]%
\>[38]{}\mathrel{\HSSym{=}}\;\HSCon{BranchSubArg}\;\HSVar{y}{}\<[E]%
\\
\>[B]{}\hsindent{3}{}\<[3]%
\>[3]{}\HSCon{SubRet}\;\HSCon{BranchOps}\;\HSSpecial{\HSSym{\{\mskip1.5mu} }\mathbin{\HSSym{\anonymous}} \HSSpecial{\HSSym{\mskip1.5mu\}}}\;{}\<[24]%
\>[24]{}\HSSpecial{\HSSym{\{\mskip1.5mu} }\HT{left}\;\HSVar{x}\HSSpecial{\HSSym{\mskip1.5mu\}}}\;{}\<[35]%
\>[35]{}\HSVar{r}\;{}\<[38]%
\>[38]{}\mathrel{\HSSym{=}}\;\HSCon{SubRet}\;\HSCon{O}\;\HSVar{r}{}\<[E]%
\\
\>[B]{}\hsindent{3}{}\<[3]%
\>[3]{}\HSCon{SubRet}\;\HSCon{BranchOps}\;\HSSpecial{\HSSym{\{\mskip1.5mu} }\mathbin{\HSSym{\anonymous}} \HSSpecial{\HSSym{\mskip1.5mu\}}}\;{}\<[24]%
\>[24]{}\HSSpecial{\HSSym{\{\mskip1.5mu} }\HT{right}\;\HSVar{y}\HSSpecial{\HSSym{\mskip1.5mu\}}}\;{}\<[35]%
\>[35]{}\HSVar{r}\;{}\<[38]%
\>[38]{}\mathrel{\HSSym{=}}\;\HSCon{BranchSubRet}\;\HSVar{r}{}\<[E]%
\\[\blanklineskip]%
\>[B]{}\hsindent{3}{}\<[3]%
\>[3]{}\HSVar{unextend}\;\mathbin{\HSCon{:}}\;\HSSym{\forall}\;\HSSpecial{\HSSym{\{\mskip1.5mu} }\HSCon{A}\HSSpecial{\HSSym{\mskip1.5mu\}}}\;\mathrel{\HSSym{\to}} \;\HSCon{AST}\;\HSCon{BranchOps}\;\HSCon{A}\;\mathrel{\HSSym{\to}} \;\HSCon{AST}\;\HSCon{O}\;\HSCon{A}{}\<[E]%
\\
\>[B]{}\hsindent{3}{}\<[3]%
\>[3]{}\HSVar{unextend}\;\mathrel{\HSSym{=}}\;\HSVar{...}{}\<[E]%
\ColumnHook
\end{hscode}\resethooks
  \caption{Extending \ensuremath{\HSCon{ASTOps}} with branching commands}
  \label{fig:ast-op-branch}
\end{figure}

Figure~\ref{fig:ast-op-branch} shows how we model these branching commands,
similar to how we model effectful commands (\Cref{sec:ast}).
Datatype \ensuremath{\HSCon{BranchCmd}} enumerates the branching commands (see module
\ensuremath{\HSCon{Dijkstra.AST.Branching}}); so far, we support commands for Booleans (\ensuremath{\HSCon{BCif}}),
coproducts (\ensuremath{\HSCon{BCeither}}), and the \ensuremath{\HSCon{Maybe}} type (\ensuremath{\HSCon{BCmaybe}}).
Each constructor of \ensuremath{\HSCon{BranchCmd}} takes the scrutinee (the subject of case
analysis) for the operation it models.
For \ensuremath{\HSCon{BranchSubArg}}, the arity of the family of subcomputations
for a branching command is given by the type of the scrutinee (e.g.,
there are two subcomputations for \ensuremath{\HSVar{if}}, so its arity is given by the two-element
type \ensuremath{\HSCon{Bool}}).
Finally, for \ensuremath{\HSCon{BranchSubRet}}, the type of result values for the subcomputations is
\ensuremath{\HSCon{A}}, the type of result values for the entire computation.

\ensuremath{\HSCon{BranchOps}} extends a given set of operations \ensuremath{\HSCon{O}\;\mathbin{\HSCon{:}}\;\HSCon{ASTOps}} with the
branching commands just described.
We take the set of command codes to the disjoint union of the command codes of
\ensuremath{\HSCon{O}} and \ensuremath{\HSCon{BranchOps}}, and extend the \ensuremath{\HSCon{SubArg}} and \ensuremath{\HSCon{SubRet}} fields
accordingly.
We also define the function \ensuremath{\HSVar{unextend}} to traverse a program AST and remove all
branching commands; this is used in the developments discussed next for
extending existing operational and predicate transformer semantics to the
branching operations.

\begin{figure}[t]
  \centering
  \begin{hscode}\SaveRestoreHook
\column{B}{@{}>{\hspre}l<{\hspost}@{}}%
\column{3}{@{}>{\hspre}l<{\hspost}@{}}%
\column{5}{@{}>{\hspre}l<{\hspost}@{}}%
\column{13}{@{}>{\hspre}l<{\hspost}@{}}%
\column{20}{@{}>{\hspre}l<{\hspost}@{}}%
\column{32}{@{}>{\hspre}l<{\hspost}@{}}%
\column{34}{@{}>{\hspre}l<{\hspost}@{}}%
\column{44}{@{}>{\hspre}l<{\hspost}@{}}%
\column{47}{@{}>{\hspre}l<{\hspost}@{}}%
\column{48}{@{}>{\hspre}l<{\hspost}@{}}%
\column{49}{@{}>{\hspre}l<{\hspost}@{}}%
\column{63}{@{}>{\hspre}l<{\hspost}@{}}%
\column{E}{@{}>{\hspre}l<{\hspost}@{}}%
\>[B]{}\HSKeyword{module}\;\HSCon{OpSemExtension}{}\<[E]%
\\
\>[B]{}\hsindent{3}{}\<[3]%
\>[3]{}\HSSpecial{\HSSym{\{\mskip1.5mu} }\HSCon{O}\;\mathbin{\HSCon{:}}\;\HSCon{ASTOps}\HSSpecial{\HSSym{\mskip1.5mu\}}}\;\HSSpecial{\HSSym{\{\mskip1.5mu} }\HSCon{T}\;\mathbin{\HSCon{:}}\;\HSCon{ASTTypes}\HSSpecial{\HSSym{\mskip1.5mu\}}}\;\HSSpecial{(}\HSCon{OpSem}\;\mathbin{\HSCon{:}}\;\HSCon{ASTOpSem}\;\HSCon{O}\;\HSCon{T}\HSSpecial{)}\;\HSKeyword{where}{}\<[E]%
\\[\blanklineskip]%
\>[B]{}\hsindent{3}{}\<[3]%
\>[3]{}\HSCon{BranchOpSem}\;\mathbin{\HSCon{:}}\;\HSCon{ASTOpSem}\;\HSCon{BranchOps}\;\HSCon{T}{}\<[E]%
\\
\>[B]{}\hsindent{3}{}\<[3]%
\>[3]{}\HSVar{runAST}\;\HSCon{BranchOpSem}\;\HSVar{m}\;\HSVar{i}\;\mathrel{\HSSym{=}}\;\HSVar{runAST}\;\HSCon{OpSem}\;\HSSpecial{(}\HSVar{unextend}\;\HSVar{m}\HSSpecial{)}\;\HSVar{i}{}\<[E]%
\\[\blanklineskip]%
\>[B]{}\HSKeyword{module}\;\HSCon{PredTransExtension}{}\<[E]%
\\
\>[B]{}\hsindent{3}{}\<[3]%
\>[3]{}\HSSpecial{\HSSym{\{\mskip1.5mu} }\HSCon{O}\;\mathbin{\HSCon{:}}\;\HSCon{ASTOps}\HSSpecial{\HSSym{\mskip1.5mu\}}}\;\HSSpecial{\HSSym{\{\mskip1.5mu} }\HSCon{T}\;\mathbin{\HSCon{:}}\;\HSCon{ASTTypes}\HSSpecial{\HSSym{\mskip1.5mu\}}}\;\HSSpecial{(}\HSCon{PT}\;\mathbin{\HSCon{:}}\;\HSCon{ASTPredTrans}\;\HSCon{O}\;\HSCon{T}\HSSpecial{)}\;\HSKeyword{where}{}\<[E]%
\\[\blanklineskip]%
\>[B]{}\hsindent{3}{}\<[3]%
\>[3]{}\HSCon{BranchPT}\;{}\<[13]%
\>[13]{}\mathbin{\HSCon{:}}\;\HSCon{ASTPredTrans}\;\HSCon{BranchOps}\;\HSCon{T}{}\<[E]%
\\
\>[B]{}\hsindent{3}{}\<[3]%
\>[3]{}\HSVar{returnPT}\;{}\<[13]%
\>[13]{}\HSCon{BranchPT}\;{}\<[32]%
\>[32]{}\mathrel{\HSSym{=}}\;\HSVar{returnPT}\;{}\<[44]%
\>[44]{}\HSCon{PT}{}\<[E]%
\\
\>[B]{}\hsindent{3}{}\<[3]%
\>[3]{}\HSVar{bindPT}\;{}\<[13]%
\>[13]{}\HSCon{BranchPT}\;{}\<[32]%
\>[32]{}\mathrel{\HSSym{=}}\;\HSVar{bindPT}\;{}\<[44]%
\>[44]{}\HSCon{PT}{}\<[E]%
\\
\>[B]{}\hsindent{3}{}\<[3]%
\>[3]{}\HSVar{opPT}\;{}\<[13]%
\>[13]{}\HSCon{BranchPT}\;\HSSpecial{(}\HT{left}\;\HSVar{x}\HSSpecial{)}\;{}\<[32]%
\>[32]{}\mathrel{\HSSym{=}}\;\HSVar{opPT}\;{}\<[44]%
\>[44]{}\HSCon{PT}\;\HSVar{x}{}\<[E]%
\\
\>[B]{}\hsindent{3}{}\<[3]%
\>[3]{}\HSVar{opPT}\;{}\<[13]%
\>[13]{}\HSCon{BranchPT}\;\HSSpecial{(}\HT{right}\;\HSSpecial{(}\HSCon{BCif}\;\HSVar{c}\HSSpecial{)}\HSSpecial{)}\;\HSVar{f}\;\HSVar{\mathit{P}}\;\HSVar{i}\;\mathrel{\HSSym{=}}\;{}\<[E]%
\\
\>[3]{}\hsindent{2}{}\<[5]%
\>[5]{}\HSSpecial{(}\HSVar{c}\;\mathbin{\HT{\equiv}}\;\HT{true}\;\mathrel{\HSSym{\to}} \;\HSVar{f}\;\HT{true}\;\HSVar{\mathit{P}}\;\HSVar{i}\HSSpecial{)}\;\mathbin{\HT{\times}}\;\HSSpecial{(}\HSVar{c}\;\mathbin{\HT{\equiv}}\;\HT{false}\;\mathrel{\HSSym{\to}} \;\HSVar{f}\;\HT{false}\;\HSVar{\mathit{P}}\;\HSVar{i}\HSSpecial{)}{}\<[E]%
\\
\>[B]{}\hsindent{3}{}\<[3]%
\>[3]{}\HSVar{opPT}\;{}\<[13]%
\>[13]{}\HSCon{BranchPT}\;\HSSpecial{(}\HT{right}\;\HSSpecial{(}\HSCon{BCeither}\;\HSVar{e}\HSSpecial{)}\HSSpecial{)}\;\HSVar{f}\;\HSVar{\mathit{P}}\;\HSVar{i}\;\mathrel{\HSSym{=}}\;{}\<[E]%
\\
\>[13]{}\hsindent{7}{}\<[20]%
\>[20]{}\HSSpecial{(}\HSSym{\forall}\;\HSVar{l}\;\mathrel{\HSSym{\to}} \;{}\<[34]%
\>[34]{}\HSVar{e}\;\mathbin{\HT{\equiv}}\;\HT{left}\;\HSVar{l}\;{}\<[48]%
\>[48]{}\mathrel{\HSSym{\to}} \;\HSVar{f}\;\HSSpecial{(}\HT{left}\;\HSVar{l}\HSSpecial{)}\;\HSVar{\mathit{P}}\;\HSVar{i}\HSSpecial{)}\;{}\<[E]%
\\
\>[13]{}\mathbin{\HT{\times}}\;{}\<[20]%
\>[20]{}\HSSpecial{(}\HSSym{\forall}\;\HSVar{r}\;\mathrel{\HSSym{\to}} \;\HSVar{e}\;\mathbin{\HT{\equiv}}\;\HT{right}\;\HSVar{r}\;{}\<[47]%
\>[47]{}\mathrel{\HSSym{\to}} \;\HSVar{f}\;\HSSpecial{(}\HT{right}\;\HSVar{r}\HSSpecial{)}\;\HSVar{\mathit{P}}\;\HSVar{i}\HSSpecial{)}{}\<[E]%
\\
\>[B]{}\hsindent{3}{}\<[3]%
\>[3]{}\HSVar{opPT}\;{}\<[13]%
\>[13]{}\HSCon{BranchPT}\;\HSSpecial{(}\HT{right}\;\HSSpecial{(}\HSCon{BCmaybe}\;\HSVar{mb}\HSSpecial{)}\HSSpecial{)}\;\HSVar{f}\;\HSVar{\mathit{P}}\;\HSVar{i}\;\mathrel{\HSSym{=}}\;{}\<[E]%
\\
\>[13]{}\hsindent{7}{}\<[20]%
\>[20]{}\HSSpecial{(}\HSSym{\forall}\;\HSVar{j}\;\mathrel{\HSSym{\to}} \;{}\<[34]%
\>[34]{}\HSVar{mb}\;\mathbin{\HT{\equiv}}\;\HT{just}\;\HSVar{j}\;{}\<[49]%
\>[49]{}\mathrel{\HSSym{\to}} \;\HSVar{f}\;\HSSpecial{(}\HT{just}\;\HSVar{j}\HSSpecial{)}\;\HSVar{\mathit{P}}\;\HSVar{i}\HSSpecial{)}\;{}\<[E]%
\\
\>[13]{}\mathbin{\HT{\times}}\;{}\<[20]%
\>[20]{}\HSSpecial{(}{}\<[34]%
\>[34]{}\HSVar{mb}\;\mathbin{\HT{\equiv}}\;\HT{nothing}\;{}\<[49]%
\>[49]{}\mathrel{\HSSym{\to}} \;\HSVar{f}\;\HT{nothing}\;{}\<[63]%
\>[63]{}\HSVar{\mathit{P}}\;\HSVar{i}\HSSpecial{)}{}\<[E]%
\ColumnHook
\end{hscode}\resethooks
  \caption{Operational and predicate transformer semantics for branching commands}
  \label{fig:ast-sem-branch}
\end{figure}

\subsubsection*{Semantics}
\label{sec:branching-semantics}

In Figure~\ref{fig:ast-sem-branch}, we assign
operational and predicate transformer semantics to \ensuremath{\HSCon{BranchOps}}.
The operational semantics for the extended set of commands reduces to the
operational semantics \ensuremath{\HSCon{OpSem}\;\mathbin{\HSCon{:}}\;\HSCon{ASTOpSem}\;\HSCon{O}\;\HSCon{T}} for the base set of commands via
\ensuremath{\HSVar{unextend}}, as shown by the definition of \ensuremath{\HSCon{BranchOpSem}}.
For the predicate transformer semantics, the precondition returned for the given
postcondition \ensuremath{\HSVar{\mathit{P}}} is expressed as a product of properties, where each component
of the product corresponds to a particular branch taken.
For example, for the case of \ensuremath{\HSCon{BCeither}\;\HSVar{e}} (where \ensuremath{\HSVar{e}\;\mathbin{\HSCon{:}}\;\HSCon{Either}\;\HSCon{B}\;\HSCon{C}}), the first
component of the product is for the case in which \ensuremath{\HSVar{e}} is of the form \ensuremath{\HT{left}\;\HSVar{l}}
for some \ensuremath{\HSVar{l}\;\mathbin{\HSCon{:}}\;\HSCon{B}}.
Recall from Section~\ref{sec:proof-engineering-pts} that expressing the proof
obligation in this way means that: the proof state is often much more
comprehensible; the proof does not need to recapitulate (for example, using
\ensuremath{\HSKeyword{with}}) the case distinction performed by the code being verified; and the
immediate next step in the proof effort is entirely type directed (copattern
matching generates the two subobligations).

\subsection{Semantic Agreement for Branching Operations}
\label{sec:branching-agree}

So far, we have augmented an arbitrary set of effectful operations with
branching commands and extended the operational and predicate transformer
semantics accordingly.
In this section, we describe our result showing that, if the original operational
and predicate transformer semantics agree---and furthermore the PTS satisfies a
certain monotonicity property, then the two extended semantics for branching
commands agree.
This ensures that the extension can be used to verify effectful code with
branching.
Furthermore, the monotonicity property is valuable in its own right when it is
easier to prove that the precondition for a postcondition that is stronger than
the one required in a given context holds.

\begin{figure}[t]
  \begin{hscode}\SaveRestoreHook
\column{B}{@{}>{\hspre}l<{\hspost}@{}}%
\column{3}{@{}>{\hspre}l<{\hspost}@{}}%
\column{5}{@{}>{\hspre}l<{\hspost}@{}}%
\column{10}{@{}>{\hspre}l<{\hspost}@{}}%
\column{19}{@{}>{\hspre}l<{\hspost}@{}}%
\column{22}{@{}>{\hspre}l<{\hspost}@{}}%
\column{33}{@{}>{\hspre}l<{\hspost}@{}}%
\column{E}{@{}>{\hspre}l<{\hspost}@{}}%
\>[B]{}\HSKeyword{record}\;\HSCon{ASTTypes}\;\mathbin{\HSCon{:}}\;\HSCon{Set}\;\HSKeyword{where}{}\<[E]%
\\
\>[B]{}\hsindent{3}{}\<[3]%
\>[3]{}\HSVar{...}{}\<[E]%
\\
\>[B]{}\hsindent{3}{}\<[3]%
\>[3]{}\HSSym{\_\!\subseteq_o\!\_}\;\mathbin{\HSCon{:}}\;\HSSym{\forall}\;\HSSpecial{\HSSym{\{\mskip1.5mu} }\HSCon{A}\HSSpecial{\HSSym{\mskip1.5mu\}}}\;\mathrel{\HSSym{\to}} \;\HSSpecial{(}\HSVar{\mathit{P}_{1}}\;\HSVar{\mathit{P}_{2}}\;\mathbin{\HSCon{:}}\;\HSCon{Post}\;\HSCon{A}\HSSpecial{)}\;\mathrel{\HSSym{\to}} \;\HSCon{Set}{}\<[E]%
\\
\>[B]{}\hsindent{3}{}\<[3]%
\>[3]{}\HSVar{\mathit{P}_{1}}\;\mathrel{\HSSym{\subseteq_o}}\;\HSVar{\mathit{P}_{2}}\;\mathrel{\HSSym{=}}\;\HSSym{\forall}\;\HSVar{o}\;\mathrel{\HSSym{\to}} \;\HSVar{\mathit{P}_{1}}\;\HSVar{o}\;\mathrel{\HSSym{\to}} \;\HSVar{\mathit{P}_{2}}\;\HSVar{o}{}\<[E]%
\\[\blanklineskip]%
\>[B]{}\hsindent{3}{}\<[3]%
\>[3]{}\HSSym{\_\!\sqsubseteq\!\_}\;\mathbin{\HSCon{:}}\;\HSSym{\forall}\;\HSSpecial{\HSSym{\{\mskip1.5mu} }\HSCon{A}\HSSpecial{\HSSym{\mskip1.5mu\}}}\;\mathrel{\HSSym{\to}} \;\HSSpecial{(}\HSVar{\mathit{pt}_{1}}\;\HSVar{\mathit{pt}_{2}}\;\mathbin{\HSCon{:}}\;\HSCon{PredTrans}\;\HSCon{A}\HSSpecial{)}\;\mathrel{\HSSym{\to}} \;\HSCon{Set}{}\<[E]%
\\
\>[B]{}\hsindent{3}{}\<[3]%
\>[3]{}\HSVar{\mathit{pt}_{1}}\;\mathrel{\HSSym{\sqsubseteq}}\;\HSVar{\mathit{pt}_{2}}\;\mathrel{\HSSym{=}}\;\HSSym{\forall}\;\HSVar{\mathit{P}}\;\HSVar{i}\;\mathrel{\HSSym{\to}} \;\HSVar{\mathit{pt}_{1}}\;\HSVar{\mathit{P}}\;\HSVar{i}\;\mathrel{\HSSym{\to}} \;\HSVar{\mathit{pt}_{2}}\;\HSVar{\mathit{P}}\;\HSVar{i}{}\<[E]%
\\[\blanklineskip]%
\>[B]{}\hsindent{3}{}\<[3]%
\>[3]{}\HSCon{MonoPT}\;\mathbin{\HSCon{:}}\;\HSSym{\forall}\;\HSSpecial{\HSSym{\{\mskip1.5mu} }\HSCon{A}\HSSpecial{\HSSym{\mskip1.5mu\}}}\;\mathrel{\HSSym{\to}} \;\HSCon{PredTrans}\;\HSCon{A}\;\mathrel{\HSSym{\to}} \;\HSCon{Set}{}\<[E]%
\\
\>[B]{}\hsindent{3}{}\<[3]%
\>[3]{}\HSCon{MonoPT}\;\HSVar{pt}\;\mathrel{\HSSym{=}}\;\HSSym{\forall}\;\HSVar{\mathit{P}_{1}}\;\HSVar{\mathit{P}_{2}}\;\mathrel{\HSSym{\to}} \;\HSVar{\mathit{P}_{1}}\;\mathrel{\HSSym{\subseteq_o}}\;\HSVar{\mathit{P}_{2}}\;\mathrel{\HSSym{\to}} \;\HSSym{\forall}\;\HSVar{i}\;\mathrel{\HSSym{\to}} \;\HSVar{pt}\;\HSVar{\mathit{P}_{1}}\;\HSVar{i}\;\mathrel{\HSSym{\to}} \;\HSVar{pt}\;\HSVar{\mathit{P}_{2}}\;\HSVar{i}{}\<[E]%
\\[\blanklineskip]%
\>[B]{}\HSKeyword{record}\;\HSCon{ASTPredTransMono}\;\HSSpecial{\HSSym{\{\mskip1.5mu} }\HSCon{OP}\HSSpecial{\HSSym{\mskip1.5mu\}}}\;\HSSpecial{\HSSym{\{\mskip1.5mu} }\HSCon{Ty}\HSSpecial{\HSSym{\mskip1.5mu\}}}\;\HSSpecial{(}\HSCon{PT}\;\mathbin{\HSCon{:}}\;\HSCon{ASTPredTrans}\;\HSCon{OP}\;\HSCon{Ty}\HSSpecial{)}\;\mathbin{\HSCon{:}}\;\HSCon{Set}\;\HSKeyword{where}{}\<[E]%
\\
\>[B]{}\hsindent{3}{}\<[3]%
\>[3]{}\HSKeyword{field}{}\<[E]%
\\
\>[3]{}\hsindent{2}{}\<[5]%
\>[5]{}\HSVar{returnPTMono}\;{}\<[19]%
\>[19]{}\mathbin{\HSCon{:}}\;{}\<[22]%
\>[22]{}\HSSym{\forall}\;\HSSpecial{\HSSym{\{\mskip1.5mu} }\HSCon{A}\HSSpecial{\HSSym{\mskip1.5mu\}}}\;\HSSpecial{(}\HSVar{x}\;\mathbin{\HSCon{:}}\;\HSCon{A}\HSSpecial{)}\;\mathrel{\HSSym{\to}} \;\HSCon{MonoPT}\;\HSSpecial{(}\HSVar{returnPT}\;\HSVar{x}\HSSpecial{)}{}\<[E]%
\\
\>[3]{}\hsindent{2}{}\<[5]%
\>[5]{}\HSVar{bindPTMono}\;{}\<[19]%
\>[19]{}\mathbin{\HSCon{:}}\;{}\<[22]%
\>[22]{}\HSSym{\forall}\;\HSSpecial{\HSSym{\{\mskip1.5mu} }\HSCon{A}\;\HSCon{B}\HSSpecial{\HSSym{\mskip1.5mu\}}}\;\HSSpecial{(}\HSVar{\mathit{f}_{1}}\;\HSVar{\mathit{f}_{2}}\;\mathbin{\HSCon{:}}\;\HSCon{A}\;\mathrel{\HSSym{\to}} \;\HSCon{PredTrans}\;\HSCon{B}\HSSpecial{)}{}\<[E]%
\\
\>[5]{}\hsindent{5}{}\<[10]%
\>[10]{}\mathrel{\HSSym{\to}} \;\HSSpecial{(}\HSSym{\forall}\;\HSVar{x}\;\mathrel{\HSSym{\to}} \;\HSCon{MonoPT}\;\HSSpecial{(}\HSVar{\mathit{f}_{1}}\;\HSVar{x}\HSSpecial{)}\HSSpecial{)}\;\mathrel{\HSSym{\to}} \;\HSSpecial{(}\HSSym{\forall}\;\HSVar{x}\;\mathrel{\HSSym{\to}} \;\HSCon{MonoPT}\;\HSSpecial{(}\HSVar{\mathit{f}_{2}}\;\HSVar{x}\HSSpecial{)}\HSSpecial{)}{}\<[E]%
\\
\>[5]{}\hsindent{5}{}\<[10]%
\>[10]{}\mathrel{\HSSym{\to}} \;\HSSpecial{(}\HSSym{\forall}\;\HSVar{x}\;\mathrel{\HSSym{\to}} \;\HSVar{\mathit{f}_{1}}\;\HSVar{x}\;\mathrel{\HSSym{\sqsubseteq}}\;\HSVar{\mathit{f}_{2}}\;\HSVar{x}\HSSpecial{)}{}\<[E]%
\\
\>[5]{}\hsindent{5}{}\<[10]%
\>[10]{}\mathrel{\HSSym{\to}} \;\HSSym{\forall}\;\HSVar{\mathit{P}_{1}}\;\HSVar{\mathit{P}_{2}}\;\HSVar{i}\;\mathrel{\HSSym{\to}} \;\HSVar{\mathit{P}_{1}}\;\mathrel{\HSSym{\subseteq_o}}\;\HSVar{\mathit{P}_{2}}\;\mathrel{\HSSym{\to}} \;\HSVar{bindPT}\;\HSVar{\mathit{f}_{1}}\;\HSVar{i}\;\HSVar{\mathit{P}_{1}}\;\mathrel{\HSSym{\subseteq_o}}\;\HSVar{bindPT}\;\HSVar{\mathit{f}_{2}}\;\HSVar{i}\;\HSVar{\mathit{P}_{2}}{}\<[E]%
\\
\>[3]{}\hsindent{2}{}\<[5]%
\>[5]{}\HSVar{opPTMono}\;{}\<[19]%
\>[19]{}\mathbin{\HSCon{:}}\;\HSSym{\forall}\;\HSSpecial{\HSSym{\{\mskip1.5mu} }\HSCon{A}\HSSpecial{\HSSym{\mskip1.5mu\}}}\;{}\<[33]%
\>[33]{}\HSSpecial{(}\HSVar{c}\;\mathbin{\HSCon{:}}\;\HSCon{Cmd}\;\HSCon{OP}\;\HSCon{A}\HSSpecial{)}\;{}\<[E]%
\\
\>[33]{}\HSSpecial{(}\HSVar{\mathit{f}_{1}}\;\HSVar{\mathit{f}_{2}}\;\mathbin{\HSCon{:}}\;\HSSpecial{(}\HSVar{r}\;\mathbin{\HSCon{:}}\;\HSCon{SubArg}\;\HSCon{OP}\;\HSVar{c}\HSSpecial{)}\;\mathrel{\HSSym{\to}} \;\HSCon{PredTrans}\;\HSSpecial{(}\HSCon{SubRet}\;\HSCon{OP}\;\HSVar{r}\HSSpecial{)}\HSSpecial{)}{}\<[E]%
\\
\>[5]{}\hsindent{5}{}\<[10]%
\>[10]{}\mathrel{\HSSym{\to}} \;\HSSpecial{(}\HSSym{\forall}\;\HSVar{r}\;\mathrel{\HSSym{\to}} \;\HSCon{MonoPT}\;\HSSpecial{(}\HSVar{\mathit{f}_{1}}\;\HSVar{r}\HSSpecial{)}\HSSpecial{)}\;\mathrel{\HSSym{\to}} \;\HSSpecial{(}\HSSym{\forall}\;\HSVar{r}\;\mathrel{\HSSym{\to}} \;\HSCon{MonoPT}\;\HSSpecial{(}\HSVar{\mathit{f}_{2}}\;\HSVar{r}\HSSpecial{)}\HSSpecial{)}{}\<[E]%
\\
\>[5]{}\hsindent{5}{}\<[10]%
\>[10]{}\mathrel{\HSSym{\to}} \;\HSSpecial{(}\HSSym{\forall}\;\HSVar{r}\;\mathrel{\HSSym{\to}} \;\HSVar{\mathit{f}_{1}}\;\HSVar{r}\;\mathrel{\HSSym{\sqsubseteq}}\;\HSVar{\mathit{f}_{2}}\;\HSVar{r}\HSSpecial{)}{}\<[E]%
\\
\>[5]{}\hsindent{5}{}\<[10]%
\>[10]{}\mathrel{\HSSym{\to}} \;\HSSym{\forall}\;\HSVar{\mathit{P}_{1}}\;\HSVar{\mathit{P}_{2}}\;\HSVar{i}\;\mathrel{\HSSym{\to}} \;\HSVar{\mathit{P}_{1}}\;\mathrel{\HSSym{\subseteq_o}}\;\HSVar{\mathit{P}_{2}}\;\mathrel{\HSSym{\to}} \;\HSVar{opPT}\;\HSVar{c}\;\HSVar{\mathit{f}_{1}}\;\HSVar{\mathit{P}_{1}}\;\HSVar{i}\;\mathrel{\HSSym{\to}} \;\HSVar{opPT}\;\HSVar{c}\;\HSVar{\mathit{f}_{2}}\;\HSVar{\mathit{P}_{2}}\;\HSVar{i}{}\<[E]%
\\
\>[B]{}\hsindent{3}{}\<[3]%
\>[3]{}\HSVar{predTransMono}\;\mathbin{\HSCon{:}}\;\HSSym{\forall}\;\HSSpecial{\HSSym{\{\mskip1.5mu} }\HSCon{A}\HSSpecial{\HSSym{\mskip1.5mu\}}}\;\HSSpecial{(}\HSVar{m}\;\mathbin{\HSCon{:}}\;\HSCon{AST}\;\HSCon{OP}\;\HSCon{A}\HSSpecial{)}\;\mathrel{\HSSym{\to}} \;\HSCon{MonoPT}\;\HSSpecial{(}\HSVar{predTrans}\;\HSVar{m}\HSSpecial{)}{}\<[E]%
\\
\>[B]{}\hsindent{3}{}\<[3]%
\>[3]{}\HSVar{predTransMono}\;\mathrel{\HSSym{=}}\;\HSVar{...}{}\<[E]%
\ColumnHook
\end{hscode}\resethooks
  \caption{Monotonicity of PTS}
  \label{fig:ast-mono}
\end{figure}

\paragraph*{Monotonicity properties}
Figure~\ref{fig:ast-mono} gives a partial listing of
\ensuremath{\HSCon{ASTPredTransMono}}, the record that describes the monotonicity lemma required by
our framework, as well as some further definitions within the scope of the
\ensuremath{\HSCon{ASTTypes}} record, which we now describe.
The type \ensuremath{\HSVar{\mathit{P}_{1}}\;\mathrel{\HSSym{\subseteq_o}}\;\HSVar{\mathit{P}_{2}}} expresses entailment of postcondition \ensuremath{\HSVar{\mathit{P}_{2}}} by \ensuremath{\HSVar{\mathit{P}_{1}}}
on all outputs (that is, that \ensuremath{\HSVar{\mathit{P}_{1}}} is at least as strong a postcondition as
\ensuremath{\HSVar{\mathit{P}_{2}}}).
For two predicate transformers \ensuremath{\HSVar{\mathit{f}_{1}}} and \ensuremath{\HSVar{\mathit{f}_{2}}}, \ensuremath{\HSVar{\mathit{f}_{1}}\;\mathrel{\HSSym{\sqsubseteq}}\;\HSVar{\mathit{f}_{2}}} is read as
saying that \ensuremath{\HSVar{\mathit{f}_{2}}} is a \emph{refinement} of \ensuremath{\HSVar{\mathit{f}_{1}}} because, for the same
postcondition \ensuremath{\HSVar{\mathit{P}}}, we have that \ensuremath{\HSVar{\mathit{f}_{2}}} produces a precondition no stronger than
that of \ensuremath{\HSVar{\mathit{f}_{1}}} (as long as both preconditions are sufficient for proving \ensuremath{\HSVar{\mathit{P}}} holds
of some program, a weaker precondition is preferable because it is more general).
Finally, monotonicity of a predicate transformer is expressed by \ensuremath{\HSCon{MonoPT}}, where
\ensuremath{\HSCon{MonoPT}\;\HSVar{f}} says that \ensuremath{\HSVar{f}} sends stronger postconditions to stronger preconditions.

Next, we consider the types of the fields
\ensuremath{\HSVar{returnPTMono}}, \ensuremath{\HSVar{bindPTMono}}, and \ensuremath{\HSVar{opPTMono}} of record \ensuremath{\HSCon{ASTPredTransMono}},
  which the user must provide to enable proving that
the predicate transformer for any effectful computation \ensuremath{\HSVar{m}\;\mathbin{\HSCon{:}}\;\HSCon{AST}\;\HSCon{Op}\;\HSCon{A}} is monotonic.
The first, \ensuremath{\HSVar{returnPTMono}}, requires that the predicate transformers in the
family assigned to \ensuremath{\HSCon{ASTreturn}} are monotonic.
The monotonicity property for composite computations (\ensuremath{\HSVar{bindPTMono}}) says that
we can map both refinement of (families of) predicate transformers
(\ensuremath{\HSSym{\forall}\;\HSVar{x}\;\mathrel{\HSSym{\to}} \;\HSVar{\mathit{f}_{1}}\;\HSVar{x}\;\mathrel{\HSSym{\sqsubseteq}}\;\HSVar{\mathit{f}_{2}}\;\HSVar{x}}) and entailment of postconditions (\ensuremath{\HSVar{\mathit{P}_{1}}\;\mathrel{\HSSym{\subseteq_o}}\;\HSVar{\mathit{P}_{2}}}) over \ensuremath{\HSVar{bindPT}}.
Similarly, for operations, \ensuremath{\HSVar{opPTMono}} says that we can map predicate
transformer refinement and postcondition entailment over the function
\ensuremath{\HSVar{opPT}} that assigns a predicate transformer to every command \ensuremath{\HSVar{c}\;\mathbin{\HSCon{:}}\;\HSCon{Cmd}\;\HSCon{Op}\;\HSCon{A}}.

\begin{figure}[t]
  \begin{hscode}\SaveRestoreHook
\column{B}{@{}>{\hspre}l<{\hspost}@{}}%
\column{3}{@{}>{\hspre}l<{\hspost}@{}}%
\column{5}{@{}>{\hspre}l<{\hspost}@{}}%
\column{15}{@{}>{\hspre}l<{\hspost}@{}}%
\column{17}{@{}>{\hspre}l<{\hspost}@{}}%
\column{19}{@{}>{\hspre}l<{\hspost}@{}}%
\column{46}{@{}>{\hspre}l<{\hspost}@{}}%
\column{60}{@{}>{\hspre}l<{\hspost}@{}}%
\column{E}{@{}>{\hspre}l<{\hspost}@{}}%
\>[B]{}\HSKeyword{module}\;\HSCon{SufficientExtension}{}\<[E]%
\\
\>[B]{}\hsindent{3}{}\<[3]%
\>[3]{}\HSSpecial{\HSSym{\{\mskip1.5mu} }\HSCon{O}\HSSpecial{\HSSym{\mskip1.5mu\}}}\;\HSSpecial{\HSSym{\{\mskip1.5mu} }\HSCon{T}\HSSpecial{\HSSym{\mskip1.5mu\}}}\;\HSSpecial{\HSSym{\{\mskip1.5mu} }\HSCon{OS}\;\mathbin{\HSCon{:}}\;\HSCon{ASTOpSem}\;\HSCon{O}\;\HSCon{T}\HSSpecial{\HSSym{\mskip1.5mu\}}}\;\HSSpecial{\HSSym{\{\mskip1.5mu} }\HSCon{PT}\;\mathbin{\HSCon{:}}\;\HSCon{ASTPredTrans}\;\HSCon{O}\;\HSCon{T}\HSSpecial{\HSSym{\mskip1.5mu\}}}{}\<[E]%
\\
\>[B]{}\hsindent{3}{}\<[3]%
\>[3]{}\HSSpecial{(}\HSCon{M}\;\mathbin{\HSCon{:}}\;\HSCon{ASTPredTransMono}\;\HSCon{PT}\HSSpecial{)}\;\HSSpecial{(}\HSCon{S}\;\mathbin{\HSCon{:}}\;\HSCon{ASTSufficientPT}\;\HSCon{OS}\;\HSCon{PT}\HSSpecial{)}\;\HSKeyword{where}{}\<[E]%
\\[\blanklineskip]%
\>[B]{}\hsindent{3}{}\<[3]%
\>[3]{}\HSCon{BranchPTMono}\;\mathbin{\HSCon{:}}\;\HSCon{ASTPredTransMono}\;\HSCon{BranchPT}{}\<[E]%
\\
\>[B]{}\hsindent{3}{}\<[3]%
\>[3]{}\HSCon{BranchPTMono}\;\mathrel{\HSSym{=}}\;\HSVar{...}{}\<[E]%
\\[\blanklineskip]%
\>[B]{}\hsindent{3}{}\<[3]%
\>[3]{}\HSVar{unextendPT}\;\mathbin{\HSCon{:}}\;{}\<[17]%
\>[17]{}\HSSym{\forall}\;\HSSpecial{\HSSym{\{\mskip1.5mu} }\HSCon{A}\HSSpecial{\HSSym{\mskip1.5mu\}}}\;\HSSpecial{(}\HSVar{m}\;\mathbin{\HSCon{:}}\;\HSCon{AST}\;\HSCon{BranchOps}\;\HSCon{A}\HSSpecial{)}{}\<[E]%
\\
\>[3]{}\hsindent{2}{}\<[5]%
\>[5]{}\mathrel{\HSSym{\to}} \;{}\<[19]%
\>[19]{}\HSVar{predTrans}\;\HSCon{BranchPT}\;\HSVar{m}\;{}\<[46]%
\>[46]{}\mathrel{\HSSym{\sqsubseteq}}\;{}\<[60]%
\>[60]{}\HSVar{predTrans}\;\HSCon{PT}\;\HSSpecial{(}\HSVar{unextend}\;\HSVar{m}\HSSpecial{)}{}\<[E]%
\\
\>[B]{}\hsindent{3}{}\<[3]%
\>[3]{}\HSVar{unextendPT}\;\mathrel{\HSSym{=}}\;\HSVar{...}{}\<[E]%
\\[\blanklineskip]%
\>[B]{}\hsindent{3}{}\<[3]%
\>[3]{}\HSVar{extendPT}\;\mathbin{\HSCon{:}}\;{}\<[15]%
\>[15]{}\HSSym{\forall}\;\HSSpecial{\HSSym{\{\mskip1.5mu} }\HSCon{A}\HSSpecial{\HSSym{\mskip1.5mu\}}}\;\HSSpecial{(}\HSVar{m}\;\mathbin{\HSCon{:}}\;\HSCon{AST}\;\HSCon{BranchOps}\;\HSCon{A}\HSSpecial{)}{}\<[E]%
\\
\>[3]{}\hsindent{2}{}\<[5]%
\>[5]{}\mathrel{\HSSym{\to}} \;{}\<[19]%
\>[19]{}\HSVar{predTrans}\;\HSCon{PT}\;\HSSpecial{(}\HSVar{unextend}\;\HSVar{m}\HSSpecial{)}\;{}\<[46]%
\>[46]{}\mathrel{\HSSym{\sqsubseteq}}\;{}\<[60]%
\>[60]{}\HSVar{predTrans}\;\HSCon{BranchPT}\;\HSVar{m}{}\<[E]%
\\
\>[B]{}\hsindent{3}{}\<[3]%
\>[3]{}\HSVar{extendPT}\;\mathrel{\HSSym{=}}\;\HSVar{...}{}\<[E]%
\\[\blanklineskip]%
\>[B]{}\hsindent{3}{}\<[3]%
\>[3]{}\HSCon{BranchSuf}\;\mathbin{\HSCon{:}}\;\HSCon{ASTSufficientPT}\;\HSCon{BranchOpSem}\;\HSCon{BranchPT}{}\<[E]%
\\
\>[B]{}\hsindent{3}{}\<[3]%
\>[3]{}\HSCon{BranchSuf}\;\mathrel{\HSSym{=}}\;\HSVar{...}{}\<[E]%
\ColumnHook
\end{hscode}\resethooks
  \caption{Sufficiency of PTS for branching operations}
  \label{fig:ast-suff-branch}
\end{figure}

\paragraph*{Agreement for the extended PTS}
Figure~\ref{fig:ast-suff-branch} shows a sketch of the proof that the extended
PTS produces, from any program \ensuremath{\HSVar{m}\;\mathbin{\HSCon{:}}\;\HSCon{AST}\;\HSCon{BranchOps}\;\HSCon{A}} and postcondition \ensuremath{\HSVar{\mathit{P}}\;\mathbin{\HSCon{:}}\;\HSCon{Post}\;\HSCon{A}}, a precondition sufficient for proving that \ensuremath{\HSVar{\mathit{P}}} holds of the result of
running \ensuremath{\HSVar{m}} with the operational semantics (\ensuremath{\HSCon{BranchSuf}} in the figure).
We prove a similar result for \ensuremath{\HSCon{Necessary}} (not shown; see module
\ensuremath{\HSCon{Dijkstra.AST.Branching}}).
The lemmas \ensuremath{\HSVar{unextendPT}} and \ensuremath{\HSVar{extendPT}} are the workhorses of these proofs: taken
together, they state that, for every such effectful program \ensuremath{\HSVar{m}}, the predicate
transformer obtained from \ensuremath{\HSCon{BranchPT}} (Figure~\ref{fig:ast-sem-branch}) is
equivalent to the one obtained from invoking the original predicate transformer
\ensuremath{\HSCon{PT}} on the result of using \ensuremath{\HSVar{unextend}} to traverse the AST of \ensuremath{\HSVar{m}} and
remove branching commands.

\section{Conclusion and Related Work}
\label{sec:concl}

We have presented an Agda framework for modeling effectful programs and
reasoning about them with predicate transformer semantics.
Our framework gives users greater flexibility in expressing intermediate
proof obligations by using a novel (to our knowledge) GADT
definition of program ASTs that enables assigning bespoke predicate
transformers to complex operations.
Demonstrating the framework's generality, we show that
the common branching operations \ensuremath{\HSVar{if}}, \ensuremath{\HSVar{maybe}}, and \ensuremath{\HSVar{either}}, can be added
\emph{\'{a} la carte} (in the sense of
Swierstra~\cite{Swi08_Data-types-a-la-Carte}) to any existing set of effects,
with the PTS extended to cover the new operations provided that the original PTS
satisfies a monotonicity condition.

Our framework codifies and generalizes techniques used in
our work \cite{NASAFM-2022} formally verifying properties of the
\textsc{LibraBFT} consensus protocol; we have used it to prove
some properties in that context, and this has confirmed
that the framework is usable for and can significantly ease real-world verification tasks.

Our work is most closely related to that of Swierstra and
Baanen~\cite{SB19_A-Predicate-Transformer-Semantics-for-Effects} on assigning
predicate transformer semantics to effectful programs.
Like us, they represent these programs with a datatype---in their case, a free
monad~\cite{HS00_Interactive-Programs-in-DTT}---and assign to it operational and
predicate transformer semantics, proving agreement of these semantics in order
to carry out verification tasks.
Our approach differs from theirs by making the datatype of program ASTs a GADT that
has the monadic bind operation as a constructor, enabling us to avoid the use of
a lemma for decomposing proof obligations found in
\cite{SB19_A-Predicate-Transformer-Semantics-for-Effects} \S 4 for composite
computations, and an expanded notion of effectful command that enables complex 
operations to be assigned bespoke predicate transformer semantics directly.
These differences give greater control to users of the framework in managing the
intermediate proof obligations generated during verification tasks.

PTS has also been used profitably with \emph{Dijkstra
  monads}~\cite{SWSCL13_Verifying-Higher-Order-Programs-Dijkstra-Monad,Jac15_Dijkstra-and-Hoare-Monads-in-Monadic-Computationa}
which leverage the monadic nature of predicate transformers (specifically, they
arise from a \ensuremath{\HSCon{Set}}-valued continuation monad
transformer~\cite{MAAMHRT19_Dijkstra-Monads-for-All}) to combine
effectful code with a formal specification.
The dependently typed programming language F* supports Dijkstra
monads~\cite{SHKRD+16_Dependent-Types-and-Multi-Monadic-Effects-in-FStar} and
integrates these with SMT solvers for automatic reasoning about effectful code.
The main advantage of our GADT-based approach (and of the free
monad approach of \cite{SB19_A-Predicate-Transformer-Semantics-for-Effects}) is
\emph{freedom of interpretation:} the \ensuremath{\HSCon{AST}} datatype describes only the syntax
of programs, meaning multiple operational and predicate transformer semantics
may be assigned to the same program.

\paragraph{Acknowledgements:} We are grateful to Victor Miraldo and Lisandra
Silva for valuable discussions and feedback on earlier versions of this paper.

\lhsinclude{body.lhs}

\lhsinclude{appendix.lhs}

\bibliographystyle{splncs04}
\bibliography{references}
\end{document}